\documentclass[10pt]{article}
\usepackage[a4paper,margin=1in]{geometry}
\usepackage{latexsym}
\usepackage{graphicx}
\usepackage{caption} 
\usepackage{lipsum}
\usepackage{wrapfig}
\usepackage{mathtools}
\usepackage{cuted}
\newcommand{\vect}[1]{\boldsymbol{#1}}
\usepackage{amsmath}
\usepackage{graphicx}
\usepackage{caption}
\usepackage{subcaption}
\usepackage{nccmath}
\usepackage{dsfont}
\usepackage{amsfonts,amssymb,amsmath}
\usepackage[shortlabels]{enumitem}
\usepackage{hyperref}
\usepackage{xcolor}
\usepackage[ruled,vlined]{algorithm2e}
\usepackage{xurl}
\usepackage{scalerel}
\usepackage{tikz}
\usepackage{setspace}
\providecommand{\keywords}[1]
{
  \small	
  \textbf{\textit{Keywords---}} #1
}

\usepackage{scalerel}
\usepackage{tikz}
\usetikzlibrary{svg.path}

\definecolor{orcidlogocol}{HTML}{A6CE39}
\tikzset{
  orcidlogo/.pic={
    \fill[orcidlogocol] svg{M256,128c0,70.7-57.3,128-128,128C57.3,256,0,198.7,0,128C0,57.3,57.3,0,128,0C198.7,0,256,57.3,256,128z};
    \fill[white] svg{M86.3,186.2H70.9V79.1h15.4v48.4V186.2z}
                 svg{M108.9,79.1h41.6c39.6,0,57,28.3,57,53.6c0,27.5-21.5,53.6-56.8,53.6h-41.8V79.1z M124.3,172.4h24.5c34.9,0,42.9-26.5,42.9-39.7c0-21.5-13.7-39.7-43.7-39.7h-23.7V172.4z}
                 svg{M88.7,56.8c0,5.5-4.5,10.1-10.1,10.1c-5.6,0-10.1-4.6-10.1-10.1c0-5.6,4.5-10.1,10.1-10.1C84.2,46.7,88.7,51.3,88.7,56.8z};
  }
}

\newcommand\orcidicon[1]{\href{https://orcid.org/#1}{\mbox{\scalerel*{
\begin{tikzpicture}[yscale=-1,transform shape]
\pic{orcidlogo};
\end{tikzpicture}
}{|}}}}

\usepackage{hyperref}

\usepackage[normalem]{ulem}
\usepackage{cite}
\usepackage{comment}
\begin{document}

\title{ Proactive Traffic Offloading  in Dynamic Integrated Multi-Satellite Terrestrial Networks }

\author{Wiem Abderrahim \orcidicon{0000-0001-5896-2307}, Osama Amin \orcidicon{0000-0002-0026-5960}, Mohamed-Slim~Alouini \orcidicon{0000-0003-4827-1793} and   Basem Shihada \orcidicon{0000-0003-4434-4334}  \\ 
\thanks{The authors are with the  Computer, Electrical and Mathematical Sciences and Engineering (CEMSE) Divison, King Abdullah University of Science and Technology (KAUST), Thuwal 23955, Makkah Prov., Saudi Arabia (e-mail:$\{$wiem.abderrahim, osama.amin, slim.alouini, basem.shihada$\}$@kaust.edu.sa)}}

\date{}

\maketitle

\begin{abstract}
The integration between the satellite network and the terrestrial network will play a key role in the upcoming sixth-generation (6G) of mobile cellular networks thanks to the wide coverage and bandwidth offered by satellite networks. To leverage this integration, we propose a proactive traffic offloading scheme in an integrated multi-satellite terrestrial network (IMSTN) that considers the future networks' heterogeneity and predicts their variability. {Our proposed offloading scheme 
hinges on traffic prediction to answer the stringent requirements of data-rate, latency and reliability imposed by heterogeneous and coexisting services and traffic namely enhanced mobile broadband (eMBB), massive machine-type communications (mMTC) and ultra-reliable low latency communication (URLLC). However, the fulfilment of these requirements during offloading in dynamic IMSTN comes at the expense of significant energy consumption and introduces inherently supplementary latency.  Therefore, our offloading scheme  aims to balance the fundamental trade-offs first between energy consumption and the achievable data-rate and second between energy consumption and latency while meeting the respective needs of the present traffic. } Our findings prove the importance of the cooperation between the multi-satellite network and the terrestrial network conditioned by traffic prediction to enhance the performance of IMTSN in terms of latency and energy consumption. 

\end{abstract}

\keywords{
Enhanced mobile broadband (eMBB), Integrated multi-satellite terrestrial  networks (IMSTN), Massive machine type communications (mMTC) , traffic offloading,  ultra-reliable low latency communication (URLLC). }

\section{Introduction}
The upcoming sixth-generation (6G) of mobile cellular networks will witness a significant expansion in data rates and coverage to meet the unprecedented growth of users, devices, applications, and services.  The 6G emerging applications and services will extend the offered fifth-generation (5G) services considerably while imposing more challenging requirements \cite{dang2020should}. Specifically, 6G is anticipated to offer more advanced and improved forms of enhanced mobile broadband (eMBB) services, massive Machine-Type Communications (mMTC) services, and Ultra-Reliable Low-Latency Communications (URLLC) services \cite{dang2020should}. Currently, under 5G, eMBB addresses the stable connection between active devices that exchange large payloads over an extended time interval with high peak rates and moderate reliability. However, mMTC manages a large number of connected devices that are sporadically active and exchange small payloads with lower transmission rates and reliability. URLLC handles intermittent transmissions between a lower number of connected devices but with stringent requirements in terms of reliability (less than $10^{-5}$ ) and low latency (1ms). 
To evolve towards the advanced forms of these services within 6G and achieve their heterogeneous and coexisting requirements, we need to develop innovative and robust network solutions. Therefore, the terrestrial networks should integrate with various aerial, sea, and space communication systems to overcome the current limitations such as traffic congestion, limited connectivity, and restricted coverage. The integration between the multi-satellite network and the terrestrial network (IMSTN) can play a crucial role in tackling these problems by widening the current terrestrial network’s capacity thanks to the wide coverage and bandwidth provided by satellite networks. Therefore, traffic offloading in these integrated networks has drawn significant attention in research recently. {We note that traffic offloading in IMSTN is a more challenging scenario compared to edge computing because satellites are in movement; whereas edge servers are fixed and can be reached at any time by the users. Moreover, traffic offloading in IMSTN is different to traffic offloading in Unmanned Aerial Vehicles (UAV) because satellites have a predefined trajectory (i.e. the satellite orbit) and periodic visibility windows; whereas UAV's trajectory is usually designed based on the use cases and the users needs. In the following, we focus only on reporting the works related to traffic offloading in IMSTN.}

\subsection{Literature Review}
Several research efforts focused on the integrated satellite-terrestrial network (ISTN) and studied different aspects related to the traffic offloading problem \cite{ zhang2019satellite,wang2018computation, boya, li2020energy,ji2020energy,abderrahim2020latency,deng2020ultra, ye2020space}. 
ISTN utilizes the traffic offloading to maximize the number of accommodated users, and their rate under a dynamic backhaul capacity constraint \cite{boya}. In this regard, it is essential to explore different offloading locations in satellite mobile edge computing (SMEC), such as the terrestrial station, the low earth orbit (LEO) satellite, and the terrestrial gateway \cite{zhang2019satellite}. Moreover, it is also important to consider users' locations in different regions such as suburban and mountainous regions and choose the suitable performance metric such as sum rate, and energy efficiency \cite{ji2020energy}. 
Also, investigating the cost matching of traffic offloading to edge servers placed in the satellite network and the terrestrial network is crucial, as studied while considering the energy consumption and the offloading delay in \cite{wang2018computation}. Content-aware offloading/caching can play a good role in improving the ISTN offloading performance by sending the unicast content through the access points, and broadcast content via the satellite \cite{li2020energy}. Furthermore, we recently considered eMBB and URLLC data traffic in designing the ISTN system, where every kind of data traffic has its requirements \cite{abderrahim2020latency}. Commercializing the ISTN solution needs a plan for pricing mechanisms for traffic offloading from the terrestrial network to the satellite network, where the frequency resources and the data service prices are optimized to boost data offloading \cite{deng2020ultra}.

Despite the recent research efforts of traffic offloading in ISTN systems, the studies focused on one fixed satellite \cite{zhang2019satellite,wang2018computation, boya, li2020energy,ji2020energy,deng2020ultra,abderrahim2020latency}. Considering multiple satellites needs careful modeling and consideration due to the interference and temporary coverage/visibility, which affect the offloading process. Moreover, the current research overlooked the heterogeneous requirements of the services of 5G and beyond networks during offloading.  Specifically, the previously discussed studies considered only one type of traffic during offloading, except \cite{abderrahim2020latency} that takes into account eMBB and URLLC. Besides, the additional latency introduced by traffic offloading was widely considered negligible. However, the latency caused by offloading to satellites, even though substantially reduced thanks to LEO,  is still intolerable by some traffic types such as URLLC.
Furthermore, most of the existing studies neglected the extra energy consumption caused by traffic offloading, assuming that the power supply is abundant for satellites. However, uncontrolled energy consumption leads to quickly-aging satellites and compels their frequent construction and launching, which are costly, and unpractical  \cite{qi2018precoding,yang2016towards,alagoz2011energy}. Therefore, there is a need to steer the research towards traffic offloading to multiple satellite systems while considering the integration with terrestrial networks along with the different traffic requirements. It is also essential to investigate the fundamental trade-off between energy and latency, which is inherent in traffic offloading. Precisely, traffic offloading in IMSTN extends the terrestrial network capacity and helps the transmissions to occur successfully with minimum latency, yet at the expense of additional power usage by the satellites. Therefore, it is necessary to balance the energy consumption and latency during traffic offloading in IMSTN.

\subsection{Contributions}

To deal with the heterogeneous service requirements and handle the energy consumption and latency trade-off, we advocate that traffic offloading in IMSTN  be proactive. Precisely, we propose employing traffic prediction as a powerful proactive scheme for traffic offloading in IMSTN. Although traffic prediction is a widely covered topic, most literature considers only the current traffic arrivals during offloading and neglects the expected future arrivals. Moreover, most of the existing works about traffic prediction, whether in terrestrial networks \cite{ar1,4,7,2} or in satellite networks \cite{8,9,12}, tend to focus on improving the prediction accuracy of the proposed algorithms. However, they rule out conducting follow-up studies that adapt the offloading problem to the predicted traffic. Therefore, in this paper, we design an innovative task offloading scheme in dynamic IMSTN based on traffic prediction. In this scheme, the arrival of the mMTC traffic is predicted, then the resources are pre-allocated to enhance the performance of offloading in IMSTN. {To the best of our knowledge, our proactive offloading scheme is the first to meet the heterogeneous requirements in terms of latency, reliability and throughput of three different  types of traffic, namely eMBB, mMTC and URLLC in a dynamic multi-satellite system}.   The main contributions of this paper can be summarized as follows: 
\begin{itemize}
    \item We consider a dynamic multi-satellite system  that cooperates with the terrestrial base stations to cover multiple users with heterogeneous traffic requirements in terms of reliability, latency and throughput. {A main dynamic aspect in our system is satellites mobility. Precisely, we assume that the satellites are  in  movement during their orbital period and can be located in different orbits. Another fundamental dynamic aspect in our system is the traffic model of three heterogeneous traffic types. Given these dynamics, our proposed offloading scheme steers smartly and proactively each traffic type to the most suitable network backhaul(s) that answers its needs in terms of latency, reliability and throughput.}  
    
    \item {We define our utility function as an energy-aware metric that seeks to establish a trade-off between the achievable data-rate of the terrestrial base stations and the total energy consumption of the satellites} with different weight factors that can be configured according to the settings and the needs of the IMSTN system. To maximize this energy-aware metric, we formulate a stochastic optimization problem that takes into account the maximum power and the maximum capacity of the satellites among the constraints along with the reliability and latency requirements of mMTC and URLLC respectively. {Then, we solve this problem by using the Lyapunov optimization framework; which yields a solution that balances a second trade-off between energy consumption and latency while satisfying the operational requirements of the different traffic types}. 
    
    \item Our simulation results show that the proposed offloading scheme  {succeeds to optimize not only the trade-off between energy consumption and achievable data-rate but also the trade-off between energy consumption and latency} while satisfying the different requirements of 5G services in terms of latency and reliability. Moreover,  our results underline the key role of traffic prediction; on which our proposed offloading scheme hinges to reduce the experienced latency in the terrestrial backhaul while conserving the same consumed power by the satellites and the same achieved data-rate for the terrestrial base stations.  

\end{itemize}

The rest of this paper is organized as follows: In section II,  our system model is presented. In section III, our problem is formulated. In section IV, the Lyapunov optimization is adopted to solve the formulated problem. In section V, our results are discussed. Finally, the paper is concluded in Section VI.

\section{System Model}

Our system model consists of a satellite network and a terrestrial network as depicted in Fig.~ \ref{fig:model}.  The terrestrial network consists of a set $\mathcal{MBS}=\{{mbs}_{1},{mbs}_{2},{\dots},{mbs}_{N}\}$ of $N$ macro base stations and a set $\mathcal{BS}=\{{bs}_{1},{bs}_{2},{\dots},{bs}_{I}\}$ of $I$ small base stations such that  each small base station ${bs}_i$ serves a set  $\mathcal{U}$ of U UEs. We consider the dense network context and the small base stations ${bs}_{i\in\{1..I\}}$ are assumed to be geographically distributed in cell $\mathcal{C}_{n\in\{1..N\}}$ according to a poisson point process (PPP).  Each small base station ${bs}_i$ in cell $\mathcal{C}_n$ is connected to a macro base station ${mbs}_n$ over a link characterized with a transmission capacity $C_i^\text{Ter}$. {The satellite network consists of a set $\mathcal{S}=\{{s}_{1},{s}_{2},{\dots},{s}_{M}\}$ of $M$ LEO satellites with a beam capacity   $C^{\text{Sat}}$.  In the scope of this paper, we consider the downlink in the satellite network {because any uncontrolled power consumption in the satellites yields to their frequent maintenance and launch; which are both costly and unpractical \cite{qi2018precoding,yang2016towards,alagoz2011energy}}. 
\begin{figure}[t]
    \centering
    \includegraphics[width=3.5in]{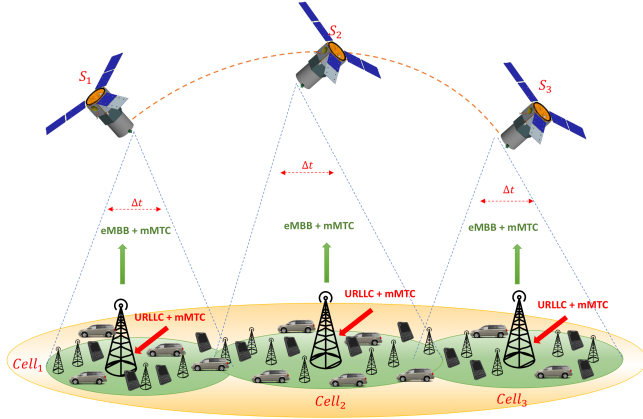}
    \caption{System Model of Traffic Prediction in IMSTN.}
    \label{fig:model}
\end{figure}
The satellite is characterized by an orbital period $T_S$, which is the time necessary to complete a full orbit around the Earth  and a service time $\Delta t$, which represents the satellite's visibility window from a fixed point on Earth.  Let us decompose the satellite's orbital period $T_S$  into $F$ frames ${\tau}_{f\in\{1..F\}}$  of duration $\Delta t$ , such that $T_S=F\times\Delta t$.   Each frame ${\tau}_{f}$ is decomposed into  $K$ time slots $t^{k \in \{1..K\}}_{f\in \{1..F\}}$ of duration $\delta$ such that  {$\Delta t=K\times \delta$} as detailed in Fig.~\ref{fig:OP}. For simplicity of notation, we drop $k$ and $f$ in the notation of time slots $t^{k}_{f}$ .  In the rest of the paper, we use $t$ to refer to $t^{k}_{f}$.
We assume that each satellite $s_j$ is allocated a downlink transmit power $P_j^\text{max}$ and a bandwidth $B_j^\text{max}$ during $t$ and that $M'$ satellites are overlapping during each service time  {$\Delta t$} for a fixed location on Earth. }

\subsection{Traffic Model in the Terrestrial Network}
 We consider three types of traffic namely eMBB, mMTC
and URLLC in our system generated by $\mathcal{U_S}$ the set of the
present UEs. We denote with $\mathcal{U}_e = \{1, .., U_1\}$ the set of $U_1$
eMBB users present in the network, $\mathcal{U}_m = \{1, .., U_2\}$ the set
of $U_2$ mMTC users and $\mathcal{U}_u = \{1, .., U_3\}$ the set of $U_3$ URLLC
users;  {such that $U_1 \geq U_2 \geq U_3$ given the inherent characteristics of these three types of traffic \cite{popovski20185g}. As a matter of fact eMBB is generated in larger volumes compared to mMTC because eMBB applications exchange important data over extended time intervals. In turn, mMTC is generated in larger amounts compared to URLLC because it involves an important number of connected devices that are sporadically active. However, URLLC  is always very limited compared to eMBB and mMTC traffic types given the very low number of connections involved during the transmission} \cite{popovski20185g,tang2019service}. Let us assume that the UE sends either eMBB traffic, mMTC traffic or URLLC traffic
such that $\mathcal{U_S} = \mathcal{U}_e \cup \mathcal{U}_m \cup \mathcal{U}_u$ and $\mathcal{U}_e \cap \mathcal{U}_m \cap \mathcal{U}_u = \varnothing$.  { Indeed, one UE can can generate one type of traffic at a specific time slot/frame, while other traffic types can be generated at successive time slots/frames \cite{popovski20185g,dos2019network,bairagi2020coexistence,korrai2020joint}. Therefore, we adjust the time granularity in our system according to the number of traffic types generated simultaneously by the same UE to cover large scale network's settings. 
}
\begin{figure}[t]
    \centering
    \includegraphics[width=5in]{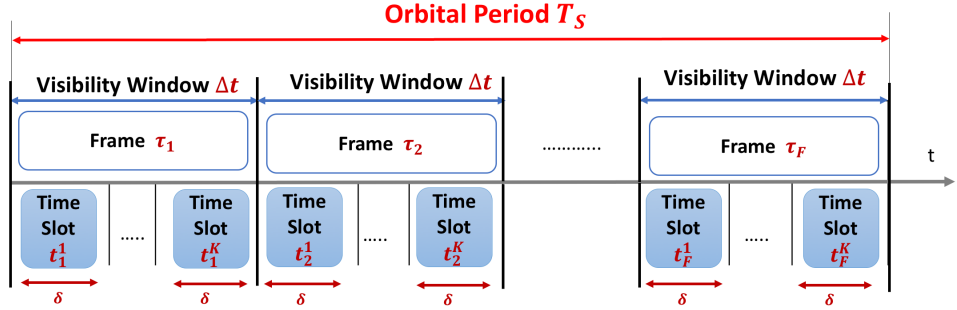}
    \caption{Timeline of the considered IMSTN.}
    \label{fig:OP}
\end{figure}
 The  eMBB traffic   is   generated   by   applications   that   exchange large  payloads  over  an  extended  time  interval \cite{popovski20185g,tang2019service}. Therefore, the inter-arrival time of eMBB  can be modeled with an exponential distribution whose events occur continuously and independently. Hence, we assume that the incoming eMBB traffic to base
station $bs_{i \in \{1..I\}}$ during time slot $t$ denoted with $A_{i,e}(t)$ follows a Poisson process with an arrival rate $\lambda^e_i$ \cite{anand}.
 However, the URLLC traffic  is  generated  by  applications  with  intermittent transmissions that exchange less important payload during short periods of time \cite{popovski20185g,tang2019service}. Therefore, we assume that the incoming URLLC traffic to base station $bs_{i \in \{1..I\}}$ during time slot $t$ denoted with $A_{i,u}(t)$ follows a Pareto distribution with an arrival rate $\lambda^u_i$ and parameters $a_i(t)$ and $x_i(t)$  \cite{anand,bennis2018ultra}. In addition, an amount $A_{i,m}(t) \leq A^\text{max}$ of mMTC
traffic arrives to base station $bs_i$; such that $A^\text{max}$ is a positive constant and $\mathop{\mathbb{E}}[A_{i,m}(t)] = \lambda^m_i$ \cite{gao2019pora}. 
 {Each base station $bs_i$ is equipped with a learning module customized to IMSTN  that can predict the future mMTC traffic by leveraging the spatio-temporal correlation between the neighboring base stations and between the subsequent satellites. The traffic in IMSTN is predicted within a prediction window of size $W_i$, (i.e., the incoming traffic in the next $W_i$ time~slots) adjusted in the learning module according to the desired requirements of the traffic. 
}

\subsection{Instantaneous Satellite's Visibility} 
  \ {To determine the satellite visibility at a given point on Earth, we establish first the time-variant coordinates of a satellite position on its orbits. The coordinates of satellite $s_j$ are the latitude $\Phi_j$ and the longitude $\Lambda_j$. 
 $\Phi_j$ can be expressed as
 $\Phi_j(t)= \arcsin(\sin(\iota)\:\sin(\theta_j(t)+\xi_0))$;
 where $\iota$ is the inclination angle of the orbital plane; $\theta_j$ is the true anomaly of satellite $s_j$ and $\xi_0$ is the argument of the perigee \cite{lutz2012satellite}.} $\Lambda_j$ can be expressed as $  \Lambda_j(t)=\Lambda_0(t)+\arccos\left(\dfrac{\cos(\theta_j(t)+\xi_0)}{\cos(\Phi_j(t))}\right)-\frac{2\pi}{T_S}\:(t-t_0)$; where $\Lambda_0$ is the longitude of the ascending node  and $t_0$ is the
time when the satellite passes the ascending node \cite{lutz2012satellite}. Then, we derive the geometric relations between satellite $s_j$ and small base station ${bs}_i$. We assume that each small base station can offload its traffic to one satellite at  time slot $t$. To this end,  this satellite should be visible by the small base station ${bs}_i$. Let us denote with {$\alpha_{\mathit{ij}}$}  the satellite's visibility variable, such that 
$
{\alpha}_{\mathit{ij}}(t)=\left\{\begin{matrix} 1 & \text{if\;} s_{j} \;\text{is} \;\text{visible}\;\text{by}\; bs_{i}\; \text{at\;} t \\
0 &
\text{otherwise}\end{matrix}\right..\label{eq:2}
$
\begin{figure}[h]
    \centering
    \includegraphics[scale=0.3]{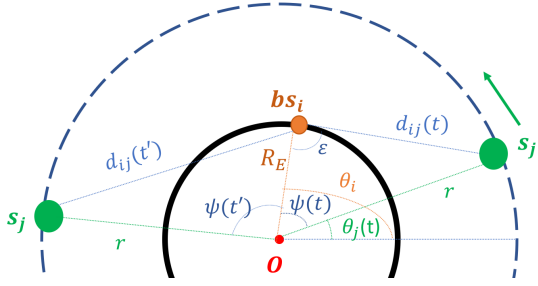}
    \caption{Satellite-Earth Geometry}
    \label{fig:my_label}
\end{figure}
\\
\noindent Hence, we define ${\alpha}_{\mathit{ij}}(t)$ as follows
$
{\alpha}_{\mathit{ij}}(t)=\!\!\left\{\begin{matrix} 1 & \text{\:if\:}d_{ij}(t) \leq d_\text{max}\;; \quad \forall \:t \in [0,T_S] \\
0 &
\text{otherwise}\end{matrix}\right.\!\!,\label{eq:3}
$
such that $d_{ij}$  is the distance (also called slant range) between the satellite ${s}_j$ and the small base station ${bs}_i$ and $d_\text{max}$ is the slant range corresponding to the minimum elevation angle. 

\ {To simplify the calculation of the satellite-Earth geometries, we assume that the  satellite's orbit is circular as depicted in Fig.~\ref{fig:my_label}   \cite{curtis2013orbital,lutz2012satellite}.} Therefore,  the satellite's true anomaly $\theta_j$ can be expressed as 
$\theta_j(t) =\frac{2\:\pi\:t}{T_S}; \quad \forall \:t \in [0,T_S]. \label{eq:3}$ \cite{lutz2012satellite}. Accordingly, the distance between the satellite ${s}_j$ and the small base station ${bs}_i$  is given by 
$
      d_{ij}(t)=\sqrt{R_E^2+r^2-2\:R_E\:r\cos\left(\theta_j(t)-\theta_i\right)} \:; \quad \forall \:t \in [0,T_S] \label{eq:4}
$
where $R_E$ is the Earth's radius, $r$ is the distance between the satellite and the Earth's center, and $\theta_i$ is the polar angle of  the small base station ${bs}_i$, respectively \cite{lutz2012satellite}. Hence, the visibility variable ${\alpha}_{\mathit{ij}}$ can be expressed for time slot $t$ as 
\begin{equation}
{\alpha}_{\mathit{ij}}(t)=\left\{\begin{matrix} 1 & \text{\:if\;}\cos\left(\frac{2\:\pi\:t}{T_S}-\theta_i\right) \geq \frac{R_E^2+r^2-d^2_\text{max}}{2\:R_E\:r} \\
0 &
\text{otherwise.}\end{matrix}\right.\!\!. \label{eq:7}
\end{equation}

\subsection{Satellite Downlink Data Rate}
During $T_S$, each satellite $s_j$ is associated a total bandwidth $W_j$ and a total power $P_j$.  { The satellite's total bandwidth and total power are fixed by the satellite manufacturers and depend on the satellite constellation characteristics. The total power and total bandwidth allocation are obtained through the link budget and depend on several factors such as the selected band, the available bandwidth, the desired coverage and the achievable throughput ... \cite{del2019technical}. Each satellite distributes its total bandwidth $W_j$ and total power $P_j$ to its associated small base stations such that ${B}_{ij}(t)$ is the bandwidth dedicated by  satellite $s_j$ to the small base station ${bs}_i$ during  $t$ and ${P}_{ij}(t)$ is the power allocated by satellite $s_j$ to the small base station ${bs}_i$ during time  $t$.  }

Therefore, the signal to noise ratio (SNR) that characterizes 
the link channel between satellite $s_j$  and  small base station ${bs}_i$ is given by
    ${SNR}_{ij}(t)\!=\!\frac{{P}_{ij}(t)\; {h}_{ij}(t)}{{N}_{0}^s\;{B}_{ij}(t)+\sum
_{j'{\in}\left\{1,..,M\right\}/j}{{P}_{ij'}(t)\;{h}_{ij'}(t)}}$, 
$\forall t \in [0,T_S]$, where  ${h}_{ij}(t)$ is the channel gain between satellite $s_j$ and small base station ${bs}_i$ during time  $t$,  ${N}_{0}^s$ is the noise power density in the space and $\sum
_{j'{\in}\left\{1,..,M\right\}/j}{{P}_{ij'}(t)\;{h}_{ij'}(t)}$ is the interference caused by the overlapping satellites.  Thus, the achievable data rate {; given by Shannon capacity formula;
${R}_{\mathit{ij}}(t)={B}_{ij}(t)\;{\log
}_{2}\left(1+{SNR}_{ij}(t)\right)$
 can be re-written as \cite{deng2020ultra,boya,li2020energy,abdu2021flexible}: 
\begin{equation}
\begin{split}
&{R}_{\mathit{ij}}(t)={B}_{ij}(t)  {\log
}_{2}\left(1+\frac{{\alpha}_{\mathit{ij}}(t)\:{P}_{ij}(t)\; \left(\frac{C}{4\:\pi\:f_c\:d_{ij}(t)}\right)^2}{{N}_{0}^S\;{B}_{ij}(t)+\sum
_{j'{\in}\left\{1,..,M\right\}/j}{{P}_{ij'}(t)\;{h}_{ij'}(t)}}\right).
\end{split}
\end{equation}}
\vspace{-1cm}

\begin{figure}[t]
    \centering
    \includegraphics[scale=0.25]{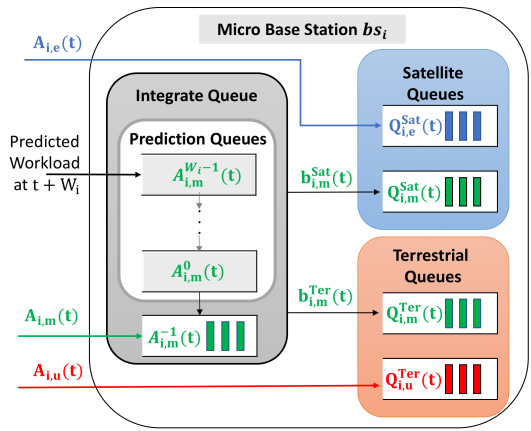}
    \caption{Queue model of the small base stations}
    \label{fig:queue}
\end{figure}

\subsection{Queuing Model}
In each small base station ${bs}_{i}$, there are different types of queues  as depicted in Fig.~\ref{fig:queue} \cite{huang2015backpressure,gao2019pora}:
\begin{enumerate}[a)] 
    \item Prediction queues with backlogs  $A_{i,m}^0(t),A_{i,m}^1(t),..,A_{i,m}^{W_i-1}(t) $, which buffer the traffic predicted to arrive in the next $W_i$ time slots.
    \item An arrival queue $A_{i,m}^{-1}(t)$ that buffers the mMTC traffic arriving in time slot $t$.
    \item Terrestrial queues $Q_{i,\kappa}^\text{Ter}$, where $\kappa \in \{\mathrm{m,u}\}$, which buffers the traffic of type $\kappa$ that will be offloaded to the terrestrial backhaul. When $\kappa=\mathrm{m}$, the traffic type is mMTC and when $\kappa=\mathrm{u}$, the traffic type is URLLC.
    \item Satellite queues $Q_{i,\kappa}^\text{Sat}$,which buffers the traffic of type $\kappa$ that will be offloaded to the satellite, where $\kappa \in \{\mathrm{m,e}\}$. When $\kappa=\mathrm{m}$, the traffic type is mMTC and when $\kappa=\mathrm{e}$, the traffic type is eMBB.
\end{enumerate}

In the following, we detail how these different queues evolve in time.

\subsubsection{Prediction Queues and Arrival Queue}
At time slot $t$, the small base station ${bs}_i$ can offload the current mMTC traffic as well as the predicted mMTC traffic within $W_i$ time slots to the terrestrial queue and to the satellite queue. 
We denote with $\mu_{i,m}^w(t)$ the output workload from $A_{i,m}^w(t), \forall\; w\; \in \{-1, 0,..., (W_{i}-1)\}$, with $b_{i,m}^\text{Ter}(t)$ the
mMTC workload offloaded to the terrestrial queue and with $b_{i,m}^\text{Sat}(t)$ the mMTC workload offloaded to the satellite queue. We note that 
    $b_{i,m}^{\nu}(t) \leq b_{i,\text{max}}^{\nu}, \forall\; \nu \in \{\text{Ter, Sat}\}$,
 where $b_{i,max}^{\nu}$ is a positive constant. 
Consequently, the sum of the output workload of all the prediction queues and the arrival queue should be equal to the sum of the workloads offloaded to the terrestrial queue and to the satellite queue respectively:$\sum_{w=-1}^{W_{i}-1} \mu_{i,m}^w(t)=b_{i,m}^\text{Ter}(t)+b_{i,m}^\text{Sat}(t)$. Hence, the prediction queues $A_{i,m}^w(t) , \forall\; w\; \in \{ 0,..., (W_{i}-1)\}$ and the arrival queue $A_{i,m}^{-1}(t)$ evolve according to the following dynamics \cite{gao2019pora,huang2015backpressure}:
\begin{enumerate}[i)]
    \item if $w=W_{i}-1$ then
    $
        A_{i,m}^{(W_{i}-1)}(t+1)=A_{i,m}(t+W_i);
    $
    such that $A_{i,m}(t+W_i)$ is the traffic predicted in time slot $t+W_i$.
    \item if $0\leq w \leq W_{i}-2$ then
    $
        A_{i,m}^{w}(t+1)=[A_{i,m}^{(w+1)}(t)-\mu_i^{(w+1)}(t)]^+;
    $
    such that $[x]^+=\max\{x,0\}.$
    \item if $w=-1$ then
    $
        A_{i,m}^{-1}(t+1)=[A_{i,m}^{-1}(t)-\mu_i^{-1}(t)]^+ +[A_{i,m}^{0}(t)-\mu_i^{0}(t)]^+;
    $
    since $A_{i,m}^{-1}$ buffers the actual traffic arriving to small base station ${bs}_i$
\end{enumerate}

Next, we introduce the integrate queue $Q_{i,m}^\text{sum}$, which is the sum of all the queues and is defined as 
$
    Q_{i,m}^\text{sum}(t)=\sum_{w=-1}^{W_{i}-1} A_{i,m}^w(t).
$
Under fully efficient service policy \cite{gao2019pora,huang2015backpressure}, $Q_{i,m}^\text{sum}$ evolves according to:
\begin{align}
 \label{eq:18'}
    &Q_{i,m}^\text{sum}(t+1)=\sum_{w=-1}^{W_{i}-1} A_{i,m}^w(t+1)\\
    &=\bigg[A_{i,m}^{-1}(t)-\mu_{i,m}^{-1}(t)\bigg]^++\bigg[A_{i,m}^{0}(t)-\mu_{i,m}^{0}(t)\bigg]^++\sum_{w=0}^{W_{i}-2}\bigg[A_{i,m}^{(w+1)}(t)-\mu_{i,m}^{(w+1)}(t)\bigg]^++A_{i,m}(t+W_i)\nonumber\\
   & =\bigg[\sum_{w=-1}^{W_{i}-1} A_{i,m}^w(t)-\sum_{w=-1}^{W_{i}-1}\mu_{i,m}^w(t)\bigg]^++A_{i,m}(t+W_i)\nonumber=\bigg[Q_{i,m}^\text{sum}(t)-\bigg(b_{i,m}^\text{Ter}(t)+b_{i,m}^\text{Sat}(t)\bigg)\bigg]^++A_{i,m}(t+W_i)\nonumber
\end{align}
The output of the integrate queue will be offloaded to the terrestrial and satellite queues.
\subsubsection{Terrestrial Queues}
 Each small base station ${bs}_i$ in cell $\mathcal{C}_n$ is connected to a macro base station ${mbs}_n$ over a link characterized with a transmission capacity $C_i^\text{Ter}(t)$ at time slot $t$. On the one hand, the terrestrial queue $Q_{i,m}^\text{Ter}$ stores the mMTC workload that will be offloaded to ${mbs}_n$ over a link capacity $C_i^\text{Ter}(t)$. Hence, the queue $Q_{i,m}^\text{Ter}$ evolves as follows:
 \begin{equation}
     Q_{i,m}^\text{Ter}(t+1)= \bigg[Q_{i,m}^\text{Ter}(t)-\bigg(C_i^\text{Ter}(t)-A_{i,u}(t)\bigg)\bigg]^++b_{i,m}^\text{Ter}(t).
      \label{eq:19'}
 \end{equation}
 The actual mMTC traffic that is offloaded to the terrestrial backhaul is $\min\{Q_{i,m}^\text{Ter}(t),C_i^\text{Ter}(t)\}$.
 
 On the other hand, the terrestrial queue $Q_{i,u}^\text{Ter}$ stores the URLLC workload that will be offloaded to ${mbs}_n$. Hence, the queue $Q_{i,u}^\text{Ter}$ evolves as follows:
 \begin{equation}
     Q_{i,u}^\text{Ter}(t+1)= \bigg[Q_{i,u}^\text{Ter}(t)-\bigg(C_i^\text{Ter}(t)-b_{i,m}^\text{Ter}(t)\bigg)\bigg]^++A_{i,u}(t).
     \label{eq:20'}
 \end{equation}

\subsubsection{Satellite Queues}
 {The small base station ${bs}_i$ can be covered by a satellite $s_j$ over a satellite link with a transmission data rate $R_{ij}(t)$}. On the one hand, the satellite queue $Q_{i,m}^\text{Sat}$ stores the mMTC workload that will be offloaded to  the satellites over a link capacity $R_{ij}(t)$. Hence, the queue $Q_{i,m}^\text{Sat}$ evolves as follows:
\begin{equation}
    Q_{i,m}^\text{Sat}(t+1)=\bigg[Q_{i,m}^\text{Sat}(t)-\bigg(\sum_{j=1}^M R_{ij}(t)-A_{i,e}(t)\bigg)\bigg]^++b_{i,m}^\text{Sat}(t).
    \label{eq:21'}
\end{equation}

On the other hand, the satellite queue $Q_{i,e}^\text{Sat}$ stores the eMBB workload that will be offloaded to the satellites. Hence, the queue $Q_{i,e}^\text{Sat}$ evolves as follows:
 \begin{equation}
     Q_{i,e}^\text{Sat}(t+1)= \bigg[Q_{i,e}^\text{Sat}(t)-\bigg(\sum_{j=1}^M R_{ij}(t)-b_{i,m}^\text{Sat}(t)\bigg)\bigg]^++A_{i,e}(t).
     \label{eq:22'}
 \end{equation}

\section{Heterogeneous Traffic Offloading Scheme}
We consider the offloading problem in dynamic IMSTN of three traffic types: eMBB, URLLC and mMTC, which have heterogeneous requirements.  Specifically, eMBB requires high data rates; while URLLC requires high reliability and  ultra-low latency. In-between, mMTC needs lower data rates than eMBB and has less stringent requirements in terms of reliability and latency than URLLC.  We take advantage of the satellite network that offers additional backhaul links to the terrestrial base stations to fulfill these heterogeneous requirements and maintain network stability.  To this end, we propose an innovative offloading scheme in IMSTN that steers the traffic smartly towards the network segment that fulfills its needs while considering the network dynamics such as satellite movement and traffic arrival. Specifically, eMBB traffic is offloaded to the satellite backhaul because it needs the high data rates offered by the satellites without stringent latency requirements. However,  URLLC  traffic is offloaded to the terrestrial backhaul, given its strict latency requirement.  mMTC traffic is offloaded to both backhauls because we consider different mMTC applications that prioritize either the high data rate requirement with low reliability (offloaded to satellite backhaul) or the latency requirement with high reliability (offloaded to terrestrial backhaul).   {In this section, we characterize our offloading scheme as an energy-aware model that optimizes the offloading decisions and the power allocation to meet the  traffic requirements in IMSTN. Therefore, we pose our optimization problem with an objective function that balances the trade-off between the power and the data rates  and with constraints that reflect the traffic requirements as detailed in the following sub-sections:}

\subsection{Utility Function}
 {Several metrics are important to study and evaluate the performance of communication systems in general and satellite networks in particular, such as data rate, power requirements, and energy efficiency.  Traditionally, the energy-efficiency is evaluated through the ratio of the data rate to the corresponding energy consumption.   In this paper, we propose a utility function $\eta_U$ that evaluates energy-efficiency differently by capturing the relative importance of the satellite transmission rate $R_{ij}(t)$ and the totally allocated power by each satellite $P_j(t)$ \cite{amin2015energy}.
 Specifically, we pose our utility function~$\eta_U$ as follows:
\begin{align}
  \label{eq:23}
  \eta_U(t)&=f(R_{ij}(t),P_j(t))
  =\frac{\beta}{I\times M}\sum_{j=1}^M\sum_{i=1}^I \sigma_i \frac{R_{ij}(t)}{C^\text{Sat}}-\frac{1-\beta}{I\times M}\sum_{j=1}^M \gamma_j \frac{P_j(t)}{P_j^\text{max}},
\end{align}
where $P_j(t)=\sum_{i=1}^I P_{ij}(t)$ is the totally allocated power by each satellite $s_j$ and $\beta$, $\sigma_{i}$ with $i\in\{1, \dots, I\}$, and $\gamma_{j}$ with $j\in\{1, \dots, M\}$ are weighting factors. }

 {The proposed utility function has a weighting parameter $\beta$ that allows changing the performance metric used in the objective function by changing $\beta$.  When we set $\beta =0$, maximizing our utility function becomes equivalent to minimizing the satellite power consumption. However,  when we impose $\beta =1$, the problem converts to maximizing the transmission data rate. Interestingly, a specific value of $\beta$ converts the problem to an energy-efficiency maximization problem and can be obtained easily using Dinkelbach optimization framework \cite{amin2015energy}.} 

The utility function $\eta_U$ is characterized also by 
 $\sigma_{i}\in [0,1]$; which is the weighting factor of small base station ${bs}_i$.  $\sigma_{i}$  prioritizes the small base station that has the most important arriving URLLC traffic. 
One way to define the weighting factors $\sigma_i$  is the following: 
  $\sigma_{i\in\{1, \dots, I\}} =\frac{\lambda_i^u}{\sum_{i=1}^I\lambda_i^u}.$

 The utility function $\eta_U$ is characterized also by   $\gamma_{j}\in [0,1]$; which is the power allocation weighting factor of satellite $s_j$. $\gamma_{j}$ prioritizes the satellite that has the largest visibility time  to boost its power allocation. One way to define the weighting factors   $\gamma_j$ is the following: $
    \gamma_j=\frac{\sum_{p=1}^P\sum_{k=1}^K\sum_{i=1}^I \alpha_{ij}(t)}{P\times K \times I}; \: \forall j\in\{1, \dots, M\},
$. We note that $\gamma_j$ is constant for the satellites in the same  plane type (i.e. polar plane or  inclined plane).

\subsection{Queues Stability and Traffic Requirements Constraints}
Our offloading scheme should   guarantee the different requirements of the considered traffic types, namely eMBB, URLLC and mMTC, in terms of high throughput, reliability and latency.  To this end, we introduce some operational constraints that ensure these requirements and guarantee the stability of the queues present in the network. First, our heterogeneous offloading scheme ensures the stability of the prediction queues and the arrival queue \cite{neely2010stochastic}:
 \begin{equation}
     \limsup \limits_{T\rightarrow \infty} \frac{1}{T} \sum_{t=1}^{T-1}\sum_{i=1}^{I}\mathop{\mathbb{E}}[Q_{i,m}^\text{sum}(t)] \leq \infty. \label{eq:26}
      \end{equation}
Our heterogeneous offloading scheme guarantees also the stability of the eMBB satellite offloading queue since no stringent requirements are needed in terms of high reliability or low latency:
 \begin{equation}
     \limsup \limits_{T\rightarrow \infty} \frac{1}{T} \sum_{t=1}^{T-1}\sum_{i=1}^{I}\mathop{\mathbb{E}}[Q_{i,e}^\text{Sat}(t)] \leq \infty. \label{eq:27}
      \end{equation} However, mMTC and URLLC traffic have stringent requirements in this regard. Therefore, our heterogeneous offloading scheme should guarantee the reliability and latency requirements for both traffic types:
     \begin{equation}
       \begin{split}
   &   \limsup \limits_{T\rightarrow \infty} \frac{1}{T} \sum_{t=1}^{T-1} \text{Pr}\Big(Q_{i,m}^\nu(t+1)> D_i^{m,\nu}\Big) \leq \epsilon_{m}^\nu;
   \forall \nu \in \{\text{Ter,Sat}\},
    \end{split}
   \label{eq:28}\end{equation}
      \begin{equation}
      \limsup \limits_{T\rightarrow \infty} \frac{1}{T} \sum_{t=1}^{T-1} \text{Pr}\Big(Q_{i,u}^\text{Ter}(t+1)> D_i^u\Big) \leq \epsilon_{u}; \label{eq:29}
 \end{equation} 
where  the queue lengths $D_i^{ m,\nu}$ and $D_i^u$ for mMTC and URLLC traffic respectively  are controlled by introducing the tolerable violation probabilities $\epsilon_{\kappa \in \{u,m\}}^{\nu \in \{\text{Ter,Sat}\}}<< 1$; such that:
$    \epsilon_{u}< \epsilon_{m}^\text{Ter}< \epsilon_{m}^\text{Sat}.$
\ { (\ref{eq:28})-(\ref{eq:29}) guarantee reliability since they ensure that the outage probability be below the negligible threshold $\epsilon_{\kappa \in \{u,m\}}^{\nu \in \{\text{Ter,Sat}\}}$.}

\subsection{Problem Formulation}
 {Let us define the offloaded mMTC traffic vector $\vect{b}=[b_{i,m}^\kappa(t)] \in \mathbb{R}^{I \times 1}$  during time slot~$t$; such that $\kappa \in \{\text{Ter,\:Sat}\}$ and  the power allocation {matrix} $\vect{P} ={[P_{ij}(t)]}\in \mathbb{R}^{(M \times N) }$; such that ${P}_{ij}(t)$ is the power allocated by  satellite $s_j$ to the small base station ${bs}_i$ during time slot~$t$.   
In the scope of our paper, we are interested in power allocation because it impacts significantly the satellite operation and lifetime; especially with the upcoming power-hungry applications and communication scenarios \cite{qi2018precoding,yang2016towards,alagoz2011energy}.  Therefore, we assume that the bandwidth is equally distributed between the base stations to simplify the optimization problem. It is worth noting that this assumption is practical since SpaceX uses one channel with a fixed bandwidth for each gateway in the downlink \cite{del2019technical}.} Hence, ${B}_{ij}(t)$ is considered constant in the rest of the paper.  The maximization of the averaged  utility function $\eta_U$  can be posed as a stochastic optimization problem; where $\vect{b}$ and $\vect{P}$ are allocated as follows:
 
 \begin{small}
\begin{align}
\max_{\vect{b},\vect{P}} & \quad \overline{\eta_U}= \limsup \limits_{T\rightarrow \infty} \frac{1}{T} \sum_{t=0}^{T-1}\mathop{\mathbb{E}}[\eta_U(t)]
\;\;\;\;\; \label{eq:20}  \\
 \mathrm{s.t.}  \; \;  
& C1:0\leq b_{i,m}^{\nu}(t) \leq b_{i,\text{max}}^{\nu}; \forall\; \nu \in \{\text{Ter, Sat}\}.\nonumber\\
&C2:\sum_{i=1}^{I}{{R}_{ij}(t)} \leq C^{\text{Sat}}\;;\quad{\forall\;}j{\;\in}\left\{1,..,M\right\}.\nonumber \\
  &C3:\sum_{i=1}^{I}{{P}_{ij}(t)}\leq P_j^\text{max} \;;\quad{\forall\;}j{\;\in}\left\{1,..,M\right\}. \nonumber
  \end{align}
 \begin{align}
  &C4: 0\leq \frac{P_{\mathit{ij}}(t)}{P_j^\text{max}} \leq {\alpha}_{\mathit{ij}}(t);  \;     {\forall\;}i{\;\in}\left\{1,..,I\right\};\;{\forall\;}j{\;\in}\left\{1,..,M\right\}.\nonumber \\
    &C5:\limsup \limits_{T\rightarrow \infty} \frac{1}{T} \sum_{t=1}^{T-1}\sum_{i=1}^{I}\mathop{\mathbb{E}}[Q_{i,e}^\text{Sat}(t)] \leq \infty.\nonumber\\
 &C6:\limsup \limits_{T\rightarrow \infty} \frac{1}{T} \sum_{t=1}^{T-1}\sum_{i=1}^{I}\mathop{\mathbb{E}}[Q_{i,m}^\text{sum}(t)] \leq \infty.\nonumber\\
 &C7:\limsup \limits_{T\rightarrow \infty} \frac{1}{T} \sum_{t=1}^{T-1} \text{Pr}\Big(Q_{i,m}^\nu(t+1)> D_i^{m,\nu}\Big) \leq \epsilon_{m}^\nu; \forall \nu \in \{\text{Ter,Sat}\},\forall\;i\;\in\left\{1,..,I\right\}.\nonumber\\
 &C8:\limsup \limits_{T\rightarrow \infty} \frac{1}{T} \sum_{t=1}^{T-1} \text{Pr}\Big(Q_{i,u}^\text{Ter}(t+1)> D_i^u\Big) \leq \epsilon_{u};\;{\forall\;}i{\;\in}\left\{1,..,I\right\}.\nonumber
\end{align}
 \end{small}
where (C1)  guarantees that the offloaded traffic is upper bounded by a maximum offloaded traffic determined based on the link capacity to prevent traffic dropping.
 (C2) reflects the limitation on the satellite backhaul capacity $C^{\text{Sat}}$.  (C3) guarantees  that the  power resources allocated to the base stations during the satellite orbital period are less than the total resources. (C4)
 guarantees that no resources are  allocated to ${bs}_i$, when it is not under the coverage of satellite $s_j$  during the considered time slot (i.e. if $\alpha_{ij}(t)=0$ then  $P_{ij}(t)=0$).  (C5)-(C8)  guarantee the queues stability and the traffic requirements  as discussed in the previous subsection through equations \eqref{eq:26}--\eqref{eq:29}.  
 
We notice that (C7) and (C8) have complex expressions given the probability of the queue length that should be evaluated for every time slot.   To eliminate the probability function, we use the following property that characterizes probabilistic events. Specifically, for  any  event $\omega$ in  the  set  of  all  possible outcomes $\Omega$, we have $\mathop{\mathbb{E}}[\mathds{1}_{\{\omega\}}]=\Pr(\omega)$ \cite{liu2019dynamic}. Therefore, we can simplify (C7) and (C8)  by using the expectation function and re-write them equivalently as ($\mathrm{C}7'$) and ($\mathrm{C}8'$): 

 \begin{small} \begin{align*}
&(C7') : \limsup \limits_{T\rightarrow \infty} \frac{1}{T} \sum_{t=1}^{T-1} \mathop{\mathbb{E}}[\mathds{1}_{\{Q_{i,m}^\nu(t+1)> D_i^{m,\nu}\}}] \leq \epsilon_{m}^\nu; \forall \nu \in \{\text{Ter,Sat}\}    \\
&(C8') : \limsup \limits_{T\rightarrow \infty} \frac{1}{T} \sum_{t=1}^{T-1} \mathop{\mathbb{E}}[\mathds{1}_{\{Q_{i,u}^\text{Ter}(t+1)> D_i^u\}}]  \leq \epsilon_{u}
  \end{align*} \end{small}
 Accordingly, our problem (\ref{eq:20}) can we rewritten as:
 \begin{align}
&\max_{\vect{b},\vect{P}} \quad \overline{\eta_U}= \limsup \limits_{T\rightarrow \infty} \frac{1}{T} \sum_{t=0}^{T-1}\mathop{\mathbb{E}}[\eta_U(t)]
\;\;\;\;\; \label{eq:pb21}  \\
&\mathrm{s.t.} \; \; \; \; C1 - C6,\: C7'- C8' \nonumber
\end{align}

 {Intuitively, the optimization problem (\ref{eq:pb21}) has so many complex constraints and cannot be directly solved. We also note that C5-C6 and C7'-C8'  are constraints
on time averaged variables. Hence, they can be satisfied only if
the queue backlogs are known at all time slots instantaneously, which is infeasible and unpractical. }

 \section{Offloading  Optimization  Framework}
 { In this section, we use the Lyapunov optimization framework to simplify our stochastic optimization problem (\ref{eq:pb21}) and deal with the constraints of the queues' stability \cite{neely2010stochastic}.  The Lyapunov optimization transforms the 
time-averaged constraints into instantaneous constraints and yields a queue mean-rate stable problem by going through different steps that include the definition of the virtual queues, the Lyapunov function and the Lyapunov Drift-plus-Penalty. Then, we decouple the obtained queue mean-rate stable problem into two independent sub-problems  that optimize the offloading decisions and the resource allocation respectively.  The resource allocation sub-problem is a non-concave maximization problem that we transformed to a convex problem thanks to a DC function reformulation and a Taylor series approximation.  Then, we solve the   approximated resource allocation sub-problem by using the successive convex approximation (SCA) iteratively.  In the following, we detail our optimization methodology:}

   \subsection{Lyapunov Optimization Problem}
We notice that our problem (\ref{eq:pb21}) seeks to optimize the time average of the utility function $\eta_U$ subject to some  constraints of time averages (C5), (C6), (C7') and (C8') related to the queues stability.  One typical way to solve this problem is to opt for the Lyapunov optimization framework.  The Lyapunov optimization framework is  appropriate for this type of problems because it requires information about the current network state without further knowledge about the future random events \cite{neely2010stochastic}. In the following, we detail the successive steps of the Lyapunov optimization.

 \subsubsection{The Virtual Queues}
 The  first step is to transform the time average constraints  into queue stability constraints.  We notice that (C5) and (C6) are in the desired form. However, (C7') and (C8') should be reformulated. Therefore, we should introduce the virtual queues  $Z_{i,m}^{\nu \in \{\text{Ter,Sat}\}}$ and $Z_{i,u}^\text{Ter}$ with the update equations 
$ Z_{i,m}^\nu(t+1)=[Z_{i,m}^{\nu}(t)+\mathds{1}_{\{Q_{i,m}^\nu(t+1)> D_i^{m,\nu}\}}-\epsilon_{m}^\nu]^+$ and $Z_{i,u}^\text{Ter}(t+1)=[Z_{i,u}^\text{Ter}(t)+\mathds{1}_{\{Q_{i,u}^\text{Ter}(t+1)> D_i^u\}}-\epsilon_{u}]^+
$ \cite{neely2010stochastic}.
Thanks to these virtual queues, the constraints (C7') and (C8') can be re-written equivalently as (C7'') and (C8'') respectively; such that:
 \begin{small}\begin{equation*}
 (C7'')\::\limsup \limits_{T\rightarrow \infty} \frac{1}{T} \sum_{t=0}^{T-1}\mathop{\mathbb{E}}[Z_{i,m}^{\nu}(t)] \leq \infty; \forall \nu \in \{\text{Ter,Sat}\};\;{\forall\;}i{\;\in}\left\{1,..,I\right\}
 \end{equation*}
 \begin{equation*}
 (C8''):\limsup \limits_{T\rightarrow \infty} \frac{1}{T} \sum_{t=0}^{T-1}\mathop{\mathbb{E}}[Z_{i,u}^\text{Ter}(t)]\leq \infty;\;{\forall\;}i{\;\in}\left\{1,..,I\right\} 
 \end{equation*}\end{small}
Hence, our problem (\ref{eq:pb21}) is equivalent to:

  \begin{align}
\max_{\vect{b},\vect{P}} & \quad \overline{\eta_U}= \limsup \limits_{T\rightarrow \infty} \frac{1}{T} \sum_{t=0}^{T-1}\mathop{\mathbb{E}}[\eta_U(t)]
\;\;\;\;\; \label{eq:pb37}  \\
 \mathrm{s.t.}  \; \;   &C1 - C6,\:C7'',\:C8'' \nonumber
\end{align} 
 
\subsubsection{ Lyapunov Function}
The second step is to derive the Lyapunov function $\mathcal{L}(\Theta(t))$, where     $\Theta(t)=[\Theta_1(t),..,\Theta_I(t)]^T$ is the queue backlog vector. $\Theta(t)=[Q(t),Z(t)]$ is the concatenation of all the actual queues $Q(t)=\{Q_{i,\kappa}^\nu(t)|{\forall\;}i{\;\in}\left\{1,..,I\right\},\nu \in \{\text{Sum,Ter,Sat}\},\kappa \in \{\text{e,m,u}\}\}$ and the virtual queues $Z(t)=\{Z_{i,\kappa}^\nu(t)|{\forall\;}i{\;\in}\left\{1,..,I\right\},\nu \in \{\text{Ter,Sat}\},\kappa \in \{\text{m,u}\}\}$ such that   \begin{align}
    \Theta_i(t)= [Q_{i,\kappa}^\nu(t),Z_{i,\kappa}^\nu(t)]
    =[Q_{i,m}^\text{Sum}(t),Q_{i,m}^\text{Ter}(t),Q_{i,m}^\text{Sat}(t),Q_{i,u}^\text{Ter}(t),Q_{i,e}^\text{Sat}(t), Z_{i,m}^\text{Ter}(t),Z_{i,m}^\text{Sat}(t),Z_{i,u}^\text{Ter}(t)].
  \end{align}

 We define the Lyapunov function $\mathcal{L}(\Theta(t))$  as function of the backlogs of the actual queues and the virtual queues \cite{neely2010stochastic}:
 
 \begin{equation}
     \mathcal{L}(\Theta(t))=\frac{1}{2} \sum_{i=1}^I{Q_{i,\kappa_1}^{\nu_1}(t)}^2+\frac{1}{2} \sum_{i=1}^I{Z_{i,\kappa_2}^{\nu_2}(t)}^2, 
 \end{equation}
 such that $\nu_1 \in \{\text{Sum,Ter,Sat}\},     \nu_2 \in \{\text{Ter,Sat}\},\kappa_1 \in \{\text{e,m,u}\},
     \kappa_2 \in \{\text{m,u}\}$.
 An expanded reformulation of $\mathcal{L}(\Theta(t))$ is given by: 
 \begin{small} \begin{align}
     \mathcal{L}(\Theta(t))&=\frac{1}{2} \sum_{i=1}^I{Q_{i,m}^{sum}(t)}^2+\frac{1}{2} \sum_{i=1}^I{Q_{i,m}^\text{Ter}(t)}^2+\frac{1}{2} \sum_{i=1}^I{Q_{i,m}^\text{Sat}(t)}^2+\frac{1}{2} \sum_{i=1}^I{Q_{i,u}^\text{Ter}(t)}^2+\frac{1}{2} \sum_{i=1}^I{Q_{i,e}^\text{Sat}(t)}^2\nonumber\\
     & + \sum_{i=1}^I{Z_{i,m}^\text{Ter}(t)}^2+\sum_{i=1}^I{Z_{i,m}^\text{Sat}(t)}^2+\sum_{i=1}^I{Z_{i,u}^\text{Ter}(t)}^2.
 \end{align} \end{small}
  \subsubsection{ Lyapunov Drift}
The third step is to define the Lyapunov drift $\Delta(\Theta(t))$, which represents the expected change in the Lyapunov function $\mathcal{L}(\Theta(t))$ over one time slot knowing that the current state is $\Theta(t)$, as follows $ \Delta(\Theta(t))=\mathop{\mathbb{E}}[\mathcal{L}(\Theta(t+1))
-\mathcal{L}(\Theta(t))|\Theta(t)] $ \cite{neely2010stochastic}. The Lyapunov drift $\Delta(\Theta(t))$ is used to push the Lyapunov function to a lower congestion state and to maintain the queues stable \cite{neely2010stochastic,peng2016energy}.

 \subsubsection{ Lyapunov Drift-plus-Penalty}
 In addition to the queue stability constraints, our objective function in problem (\ref{eq:pb37}) is to maximize the time average of the utility function $\eta_U$. Therefore, we define the drift-plus-penalty, whose expression is given by \cite{neely2010stochastic}:
 
 \begin{equation}
         \Delta(\Theta(t))-V\mathop{\mathbb{E}}[\eta_U(t)|\Theta(t)],
         \label{eq:drift}
 \end{equation}
 
 {where $V>0$ is a weighting factor that reflects how much we emphasize on $\eta_U$ maximization.
 When we choose $V = 0$,  we only minimize the drift; which leads to stable queues and lower latency but not necessarily to an optimal utility function. When we  choose $V > 0$,  we include the utility function in the optimization and we establish a smooth trade-off between latency reduction and utility maximization. According to Lyapunov theory, the time average of our utility function  deviates by at most $O(\frac{1}{V} )$ from optimality, with a time average queue backlog bound of $O(V)$ \cite{neely2010stochastic}. }
 Thanks to Lyapunov optimization framework, our problem (\ref{eq:pb37})
 is relaxed by introducing the trade-off parameter $V$ between the utility function and the latency. Hence, our problem (\ref{eq:pb37}) can be re-written as minimizing the drift-plus-penalty and can be posed as follows:
 \begin{small}\begin{align}
&\min_{\vect{b},\vect{P}} \quad \Delta(\Theta(t))-V\mathop{\mathbb{E}}[\eta_U(t)|\Theta(t)]\label{eq:39''}\\
&\mathrm{s.t.}  \; \; \; C1-C4  \nonumber
\end{align}\end{small}
After some simplifications detailed in Appendix A and thanks to the derivation of  the upper bound of the drift-plus-penalty, our problem (\ref{eq:39''}) can be equivalently re-written as: 
 \begin{small}\begin{align}
\label{eq:45'}
\min_{\vect{b},\vect{P}} \: 
&\frac{V(1-\beta)}{I\times M}\sum_{j=1}^M  \sum_{i=1}^I \gamma_j P_{ij}(t)-\frac{V\beta}{I \times M}\sum_{j=1}^M
\sum_{i=1}^I \sigma_i \frac{R_{ij}(t)}{C^\text{Sat}}-\sum_{i=1}^I\sum_{j=1}^M R_{ij}(t)\bigg(2Q_{i,m}^\text{Sat}(t)+Q_{i,e}^\text{Sat}(t)+Z_{i,m}^\text{Sat}(t)+ A_{i,e}(t)\bigg)\nonumber\\ 
&+\sum_{i=1}^Ib_{i,m}^\text{Ter}(t) \bigg(2Q_{i,m}^\text{Ter}(t)+2Q_{i,u}^\text{Ter}(t)+Z_{i,m}^\text{Ter}(t)+Z_{i,u}^\text{Ter}(t)+ 2A_{i,u}(t) -2C_i^\text{Ter}(t)-Q_{i,m}^\text{Sum}(t)\bigg) \\
&+\sum_{i=1}^Ib_{i,m}^\text{Sat}(t) \bigg(2Q_{i,m}^\text{Sat}(t)+Q_{i,e}^\text{Sat}(t)+Z_{i,m}^\text{Sat}(t)+ A_{i,e}(t)-Q_{i,m}^\text{Sum}(t)\bigg)\nonumber  \\ 
&\mathrm{s.t.} \quad C1-C4 \nonumber
\end{align}
\end{small}

\subsection{Offloading Problem Decomposition}
 We notice that our objective function $f(\vect{b},\vect{P})$ in problem (\ref{eq:45'}) can be written as $f(\vect{b},\vect{P})=f_1(\vect{b})+f_2(\vect{P})$; such that:
 \begin{small} \begin{flalign}
f_1(\vect{b})&=\sum_{i=1}^Ib_{i,m}^\text{Ter}(t) \bigg(2Q_{i,m}^\text{Ter}(t)+2Q_{i,u}^\text{Ter}(t)+Z_{i,m}^\text{Ter}(t)+Z_{i,u}^\text{Ter}(t)+ 2A_{i,u}(t) -2C_i^\text{Ter}(t)-Q_{i,m}^\text{Sum}(t)\bigg)\nonumber &&\\
&+  \sum_{i=1}^Ib_{i,m}^\text{Sat}(t)\bigg(2Q_{i,m}^\text{Sat}(t)+Q_{i,e}^\text{Sat}(t)
+Z_{i,m}^\text{Sat}(t)+A_{i,e}(t)-Q_{i,m}^\text{Sum}(t)\bigg)&&
 \end{flalign} \end{small}
 \begin{small}
 \begin{equation}
   f_2(\vect{P})=   \frac{V(1-\beta)}{I\times M}\sum_{j=1}^M  \sum_{i=1}^I \gamma_j P_{ij}(t)-\frac{V\beta}{I \times M}\sum_{j=1}^M
\sum_{i=1}^I \sigma_i \frac{R_{ij}(t)}{C^\text{Sat}}-\sum_{i=1}^I\sum_{j=1}^M R_{ij}(t)\bigg(2Q_{i,m}^\text{Sat}(t)+Q_{i,e}^\text{Sat}(t)+Z_{i,m}^\text{Sat}(t)+ A_{i,e}(t)\bigg).
 \end{equation}  \end{small}
 Given this reformulation of the objective function as the sum of two independent functions and since the constraint (C1) is independent of the constraints (C2)-(C4), we can decompose our problem (\ref{eq:45'}) into two independent  sub-problems related respectively to the offloading decisions of the mMTC traffic \textbf{(SP1) } and to the resource allocation  \textbf{(SP2)} as  detailed and rearranged in the next subsections \cite{boyd2004convex}.
 \subsubsection{Sub-problem 1: Offloading Decisions}\hfill\\
 In each time slot, the offloaded mMTC traffic by the small base station via the terrestrial link and the satellite link  respectively should be determined by solving the following sub-problem \textbf{(SP1)}: 
\begin{align}
\label{eq:520}
\textbf{(SP1):} \;&\min_{b}  \quad f_1(\vect{b}) \\
&\mathrm{s.t.} \quad C1:0\leq b_{i,m}^{\nu}(t) \leq b_{i,\text{max}}^{\nu}; \forall\; \nu \in \{\text{Ter, Sat}\}. \nonumber
\end{align}
 Our variable $\vect{b}=[b_{i,m}^{\kappa\in \{\text{Ter, \, Sat}\}}(t)]$ consists of the offloaded mMTC traffic during time slot~$t$ to the terrestrial backaul $b_{i,m}^\text{Ter}(t)$ and to the satellite backhaul $b_{i,m}^\text{Sat}(t)$ respectively. Since $b_{i,m}^\text{Ter}(t)$ and $b_{i,m}^\text{Sat}(t)$ are independent $\forall i \in \{1, \dots, I\}$, we can decompose \textbf{(SP1)} into  two independent sub-problems related to mMTC offloading to the terrestrial backhaul and  to the satellite backhaul respectively  as follows: 
\begin{small} \begin{flalign}
& \quad \min_{b_{i,m}^\text{Ter}}  \quad
b_{i,m}^\text{Ter}(t) \bigg(2Q_{i,m}^\text{Ter}(t)+2Q_{i,u}^\text{Ter}(t)+Z_{i,m}^\text{Ter}(t)+Z_{i,u}^\text{Ter}(t)+ 2A_{i,u}(t) -2C_i^\text{Ter}(t)-Q_{i,m}^\text{Sum}(t)\bigg)\nonumber&&\\
& \quad \mathrm{s.t.} \quad 0\leq b_{i,m}^\text{Ter}(t) \leq b_{i,\text{max}}^\text{Ter}.&&
\end{flalign}\end{small}
and

\begin{small} \begin{flalign}
& \quad \min_{b_{i,m}^\text{Sat}}  \quad b_{i,m}^\text{Sat}(t) \bigg(2Q_{i,m}^\text{Sat}(t)+Q_{i,e}^\text{Sat}(t)+Z_{i,m}^\text{Sat}(t))+ A_{i,e}(t)-Q_{i,m}^\text{Sum}(t)\bigg)&&\nonumber \\ 
&  \quad \mathrm{s.t.} \quad 0\leq b_{i,m}^\text{Sat}(t) \leq b_{i,\text{max}}^\text{Sat}.&&\label{eq:46}
\end{flalign}\end{small}

The optimized solutions $b_{i,m}^\text{Ter*}$ and $b_{i,m}^\text{Sat*}$ are given by:
\begin{small}\begin{align}   b_{i,m}^\text{Ter*}(t)=
\begin{cases}
   b_{i,\text{max}}^\text{Ter} ,& \text{if }\quad 2Q_{i,m}^\text{Ter}(t)+2Q_{i,u}^\text{Ter}(t)+Z_{i,m}^\text{Ter}(t) +Z_{i,u}^\text{Ter}(t)
 +2A_{i,u}(t)-2C_i^\text{Ter}(t)<Q_{i,m}^\text{Sum}(t)\\
    0,              & \text{otherwise}
\end{cases}
\label{eq:47'}
\end{align}\end{small}

and
\begin{small}
\begin{flalign}
b_{i,m}^\text{Sat*}(t)=
\begin{cases}
   b_{i,\text{max}}^\text{Sat} ,& \text{if } 2Q_{i,m}^\text{Sat}(t)+Q_{i,e}^\text{Sat}(t)+Z_{i,m}^\text{Sat}(t)
+ A_{i,e}(t)<Q_{i,m}^\text{Sum}(t) \quad \quad \quad \quad \quad \quad\quad \quad \quad\quad \quad \\
    0,              & \text{otherwise}
\end{cases}&
\label{eq:48'}
\end{flalign}
\end{small}
Specifically, the small base station compares the backlog size of its integrate queue to the sum of the backlogs of the queues  detailed in (\ref{eq:47'}) and (\ref{eq:48'}). When this latter queue is more loaded, the small base station offloads as much traffic as possible until reaching $b_{i,\text{max}}^\text{Ter}$ and $b_{i,\text{max}}^\text{Sat}$. Otherwise, no traffic is offloaded. The selection of the satellite queue or the terrestrial queue is imposed by the tolerable violation probabilities $\epsilon_{m}^\text{Sat}$ and $\epsilon_{m}^\text{Ter}$. These requirements are guaranteed through the presence of the virtual queues $Z_{i,m}^\text{Sat}(t)$ and $Z_{i,m}^\text{Ter}(t)$ respectively. Once the problem \textbf{(SP1)} is solved, we consider the optimized solution $b^*$ to update the integrate queue $Q_{i,m}^\text{sum}$, the terrestrial queues $Q_{i,m}^\text{Ter}$, $Q_{i,u}^\text{Ter}$ and the satellite queues $Q_{i,m}^\text{Sat}$, $Q_{i,e}^\text{Sat}$ (c.f.  (\ref{eq:18'})-(\ref{eq:22'})).

 \subsubsection{Sub-problem 2: Resource Allocation}\hfill\\
 The power resources $\vect{P}$ allocated to the small base stations by the satellites should be determined by solving the following sub-problem \textbf{(SP2)}: 
 
\begin{align}
\label{eq:521}
\textbf{(SP2):} \;&\min_{P}  \quad f_2(\vect{P})  \\
&\mathrm{s.t.} \quad C2-C4 \nonumber
\end{align}
After rearranging \textbf{(SP2)} and expressing ${R}_{ij}(t)$ as function of ${P}_{ij}(t)$, we obtain equivalently the following problem: 
\begin{small}  \begin{align}
  \label{eq:500}
\max_{\vect{P}}  &
\sum_{j=1}^M  \sum_{i=1}^I \! \Bigg(\! X_{i}{B}_{ij}(t)\!\ln \! \bigg(\! 1\!+\! \frac{h_{ij}(t){P}_{ij}(t)}{{N}_{0}{B}_{ij}(t)\!+\!\sum_{j'{\in}\left\{1,M\right\}/j}{{P}_{ij'}(t) {h}_{ij'}(t)}}\!\bigg)-X_{j}P_{ij}(t)\Bigg)  \\
 \mathrm{s.t.} \; 
   &C2:\sum_{i=1}^{I}{\!{B}_{ij}(t){\log
}_{2}\!\bigg(\!1\!+\!\frac{{P}_{ij}(t)h_{ij}(t)}{{N}_{0}{B}_{ij}(t)+\sum_{j'{\in}\left\{1,M\right\}/j}{{P}_{ij'}(t)\;{h}_{ij'}(t)}}\!\bigg)}\! \leq \! C^{\text{Sat}}\nonumber \\
 \footnotesize
 &C3:\sum_{i=1}^{I}{{P}_{ij}(t)}\leq P_j^\text{max} \;;\quad{\forall\;}j{\;\in}\left\{1,..,M\right\} \nonumber \\
 &C4: 0\leq \frac{P_{\mathit{ij}}(t)}{P_j^\text{max}}\leq {\alpha}_{\mathit{ij}}(t);  \;     {\forall\;}i{\;\in}\left\{1,..,I\right\};\;{\forall\;}j{\;\in}\left\{1,..,M\right\}\nonumber
\end{align} \end{small}
such that $X_{j}=\frac{V(1-\beta)\gamma_j}{I\: M\:P_j^\text{max}}$ and 
$X_{i}=\Big(\frac{V\beta\sigma_i}{C^\text{Sat}\:I\: M}+2Q_{i,m}^\text{Sat}(t)+Q_{i,e}^\text{Sat}(t)+Z_{i,m}^\text{Sat}(t)+ A_{i,e}(t)\Big)$.\\

Our problem (\ref{eq:500}) is non-concave because our objective function is non-concave and constraint C2 is non-convex in $\vect{P}$. To deal with this issue, we express them as the difference of two concave functions $\mathcal{F}(P)=\sum_{j=1}^M  \sum_{i=1}^I {\mathcal{F}}(P_{ij}(t))$ and  $\mathcal{G}(P)=\sum_{j=1}^M  \sum_{i=1}^I {\mathcal{G}}(P_{ij}(t))$; where
 \begin{small}
\begin{equation}
   \mathcal{F}(P_{ij}(t))=X_{i}{B}_{ij}(t)\ln\Big({\sum_{j}{{P}_{ij}(t)\;{h}_{ij}(t)+{N}_{0}\:{B}_{ij}(t)}\Big)} \label{eq:55}
\end{equation}\end{small}

\begin{small}
\begin{equation}
       \mathcal{G}(P_{ij}(t))=X_{i}{B}_{ij}(t)\ln\Big({\sum_{j'{\in}\left\{1,M\right\}/j}{{P}_{ij'}(t)\;{h}_{ij'}(t)}+{N}_{0}\:{B}_{ij}(t)\Big)} \label{eq:56}. 
\end{equation}
\end{small}
Then, we approximate $\mathcal{F}$ with $\widehat{\mathcal{F}}$ by   using the first order representation of Taylor series at $\widehat{P}$ as follows  \cite{boyd2004convex}: 
 
\begin{small}
\begin{equation}
    \mathcal{F}(P) \leq \widehat{\mathcal{F}}(P)=\mathcal{F}(\widehat{P})+ \nabla_\mathcal{F}(\widehat{P})^T(P-\widehat{P}) \\
\label{eq:650}
\end{equation}
\end{small}
where a developed form of $\widehat{\mathcal{F}}(P)$ is given by: 
\begin{small}\begin{align}
\widehat{\mathcal{F}}(P)=&\sum_{j=1}^M  \sum_{i=1}^I \widehat{\mathcal{F}}(P_{ij}(t)) \\ 
=&\sum_{j=1}^M  \sum_{i=1}^I X_{i}{B}_{ij}(t)\ln\Big({\sum_{j}{\hat{P}_{ij}(t)\;{h}_{ij}(t)+{N}_{0}\:{B}_{ij}(t)}\Big)} +\sum_{j=1}^M  \sum_{i=1}^I \frac{X_{i}{B}_{ij}(t)h_{ij}(t)({P}_{ij}(t)-\hat{P}_{ij}(t))}{\sum_{j}{\hat{P}_{ij}(t)\;{h}_{ij}(t)}+{N}_{0}\:{B}_{ij}(t)}
\label{eq:651}
\end{align}
\end{small}
Therefore, a simplified approximated problem of (\ref{eq:500}) can be reformulated by using the aforementioned first-order approximation (\ref{eq:651}) as follows: 
 
  \begin{align}
  \label{eq:670}
&\max_{\vect{P}}  \quad
\medmath{\widehat{\mathcal{F}}(P)-\mathcal{G}(P)   -\sum_{j=1}^M  \sum_{i=1}^IX_{j}\:P_{ij}(t)}  \\
\mathrm{s.t.}  \; \; \; & C2:\medmath{\sum_{i=1}^I\Bigg( \frac{\widehat{\mathcal{F}}(P_{ij}(t))}{X_i}-\mathcal{G}(P_{\mathit{ij}})\Bigg) -  C^{\text{Sat}}\leq 0;\;{\forall\:}j{\:\in}\left\{1,M\right\}}\nonumber\\
  &C3-C4 \nonumber 
\end{align}
 
 { Thanks to reformulating the objective function of \textbf{(SP2)} as well as   constraint C2 in terms of a DC function, and approximating it using first-order Taylor series, the reduced problem \eqref{eq:670} becomes a convex problem. It is worth to emphasize that this approximation does not violate the problem. However, the proposed bound reduces the feasibility space resulting in a sub-optimal solution.  To solve the reformulated problem \textbf{(SP2)}, we use  SCA and solve the approximated convex problem \eqref{eq:670} iteratively where $\widehat{P}$ is updated at each iteration using the previous solution. It is worth to mention that the choice of the initial point is critical to reduce the required number of iterations and hence to reach the sub-optimal solution quickly.} 
 \begin{algorithm}
 {
\SetAlgoLined
\SetKwInput{KwInput}{Input}
\SetKw{KwInitialize}{Initialize:}
\KwInput{$C^\text{Sat}$,$P^{\text{max}}$,$\;\alpha$,$\;h_{ij}\:$,$\;{B}_{ij}\;$,$X_i,X_j$,$(i,j)\in\{1,N\}\times\{1,M\}$}
\KwInitialize{ $\epsilon_\mathrm{error}$, $\widehat{P}$
} \\
 \While{$\left(|\widehat{P} - {P}^{*}| \geq \epsilon_{\mathrm{error}}\right)$}{
Solve problem (\ref{eq:670}) to get ${P}^{*}$;\\
 $\widehat{P}\; \gets  {P}^{*} $;
  }}
 \caption{Resource Allocation}
\end{algorithm} 

 Algorithm 1 solves the resource allocation problem by using the SCA approach. First, we start with an initial arbitrarily value for  $\widehat{P}$ and stopping error, $\epsilon_{\mathrm{error}}$, for the iterative algorithm. We can  start with  evenly distributed power between all small base stations such that $\widehat{P}=\big[\frac{P_j^\text{max}}{I}\big]_{j\in\{1,M\}}$. In each iteration of Algorithm 1, we solve the convex optimization problem (\ref{eq:670}) by using the Interior Point algorithm, where $\widehat{P}$ uses the solution of the previous iteration $P^*$.   {The computational complexity of the initialization phase is $\mathcal{O}(M\times I \times K \times F)$; where we set the initial values for our problem settings such as  the channel gain and the slant range... Moreover, the computational complexity of the resource allocation phase depends on the timescale of our system and is given by $\mathcal{O}(K \times F)$.}  {Once the problem \textbf{(SP2) } is solved, we obtain the sub-optimal  power allocated to the small base stations by the satellites.} We consider the optimized solution $P^*$  in order to estimate the rate of each base station and to update the queues $ Q_{i,m}^\text{Sat}$ and  $Q_{i,e}^\text{Sat}$ in each time slot (c.f. \eqref{eq:21'} and \eqref{eq:22'}).

 \section{Results and Discussion}

  In this section, we evaluate the performance of our proactive offloading scheme  by carrying out four  different simulations.   First, we investigate the impact of the number of satellites in service on our scheme performance. Second, we investigate how our scheme handles the first trade-off between energy consumption and achievable rate by assessing the impact of the parameter $\beta$; which was defined in our energy-aware utility function (c.f. (\ref{eq:23})). Then, we investigate how our scheme handles the second trade-off between energy consumption and latency; which was introduced by approximation of our optimization problem (\ref{eq:pb21}) through the Lyapunov optimization framework. Therefore, we assess the impact of the parameter $V$ (c.f (\ref{eq:drift})) on the offloading performance. Finally, we investigate the role of traffic prediction to improve offloading in dynamic IMSTN by assessing  the impact of the window size.   We  conduct  our simulations based on the features of the Telesat LEO constellation \cite{del2019technical}. The  simulations  parameters  are  detailed   in Table \ref{table:2} for the satellite network and in  Table  \ref{table:1} for the terrestrial network.  { Without loss of generality, we consider the large-scale fading particularly in our channel model because it is the main component that significantly affects the signal quality since satellite communications are established over long distances  \cite{zhao2020spatial}. Hence, the channel gain ${h}_{ij}$  between satellite $s_j$ and small base station ${bs}_i$  mainly depends on the path loss and  can be modeled as 
  $ h_{ij}(t) = {\alpha}_{\mathit{ij}}(t)  h_{ij}^{a}(t)   \label{eq:9}$; such that the coefficient $h_{ij}^\mathrm{a}(t)$ captures the effect of the path loss between satellite $s_j$ and small base station ${bs}_i$ and is given by  $h_{ij}^\mathrm{a}(t)=\left(\frac{C}{4\:\pi\:f_c\:d_{ij}(t)}\right)^2 \label{eq:10}$; where $C$ is the light speed, $f_c$ is the carrier frequency and $d_{ij}(t)$ is the slant range (i.e. distance) between satellite $s_j$ and small base station ${bs}_i$ \cite{zhao2020spatial}.}

  \begin{table}[hbt!]
\caption{ Telesat Constellation Features per Downlink Beam \cite{del2019technical}}
\centering
\begin{tabular}{|c |c| c |c|} 
 \hline
Parameter & Telsat  \\ [1ex] 
 \hline\hline 
 Orbital Period $T_S$  & 6627.6 s \\[1ex] 
 Number of Satellites per Polar Plane & 12 \\[1ex] 
 Number of Satellites per Inclined Plane & 9 \\[1ex] 
 Inclination for Polar Planes  & $99.5^{\circ}$ \\[1ex] 
 Inclination for Inclined Planes  & $37.4^{\circ}$ \\ [1ex] 
 Satellite Bandwidth ($B_j^\text{max}$) & 0.25 Ghz \\  [1ex] 
 Satellite Link Capacity ($C^\text{Sat}$) & 558.7 Mbps \\[1ex] 
 Satellite Transmit Power ($P_j^\text{max}$)  & 1000 W \\
 [1ex] 
 \hline 
\end{tabular}
\label{table:2}
\end{table}

 \begin{table}[hbt!]
\caption{ Terrestrial Network Parameters}
\centering
\begin{tabular}{|c |c| } 
 \hline
Parameter & Numerical Value\\ [1ex] 
 \hline\hline 
  URLLC Arrival Rate ($\lambda_u$)   & [10-100] kbps \\[1ex] 
 eMBB Arrival Rate ($\lambda_e$)   & [100-150] kbps \\[1ex] 
  Tolerable  Violation Probability ($\epsilon_u$) & $10^{-5}$ \\ [1ex] 
  Tolerable  Violation Probability ($\epsilon_m^\text{Ter}$) & $10^{-3}$  \\[1ex] 
Tolerable  Violation Probability ($\epsilon_m^\text{Sat}$) & $10^{-2}$  \\[1ex]
Maximum Queue Length ($D_\text{i}^\text{u}$)  & $0.3 \times C^\text{Ter}$ \\ [1ex] 
Maximum Queue Length ($D_\text{i}^\text{m,Ter}$)  & $0.7 \times C^\text{Ter}$ \\[1ex] 
Maximum Queue Length ($D_\text{i}^\text{m,Sat}$) & $0.5 \times C^\text{Sat}$ \\[1ex] 
Maximum Offloaded Traffic to Satellite ($ b_\text{max}^\text{Sat}$)  & ${C^\text{Sat}}$ \\[1ex]  
Maximum Offloaded Traffic to Terrestrial ($  b_\text{max}^\text{Ter}$) & $C^\text{Ter}$ \\ [1ex] 
 \hline 
\end{tabular}
\label{table:1}
\end{table} 

\newpage
\begin{figure*}
    \centering
     \begin{subfigure}[b]{0.475\textwidth}
    \includegraphics[width=3.5in]{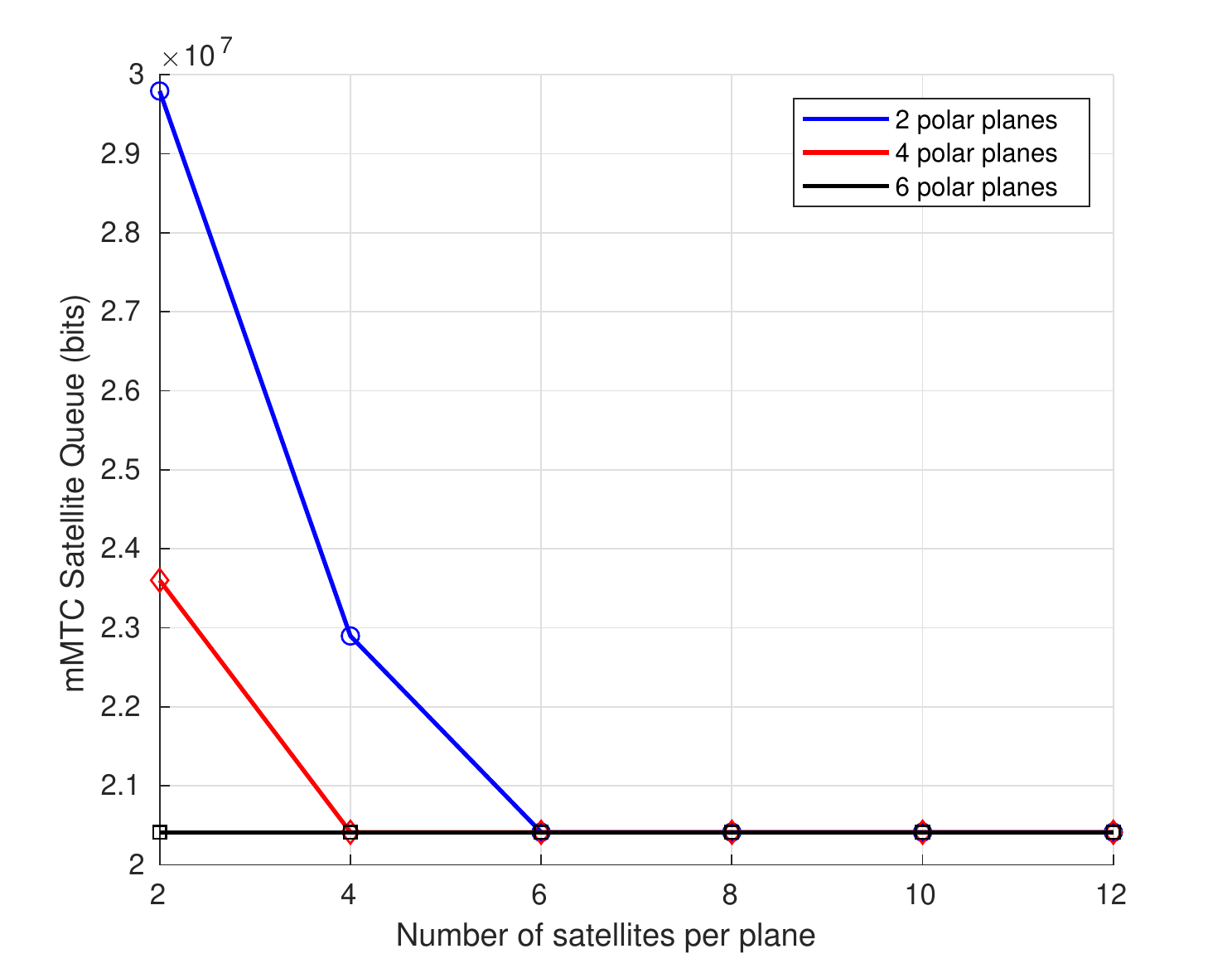}
    \caption{}
    \label{fig:16}
     \end{subfigure}
     \begin{subfigure}[b]{0.475\textwidth}
    \includegraphics[width=3.5in]{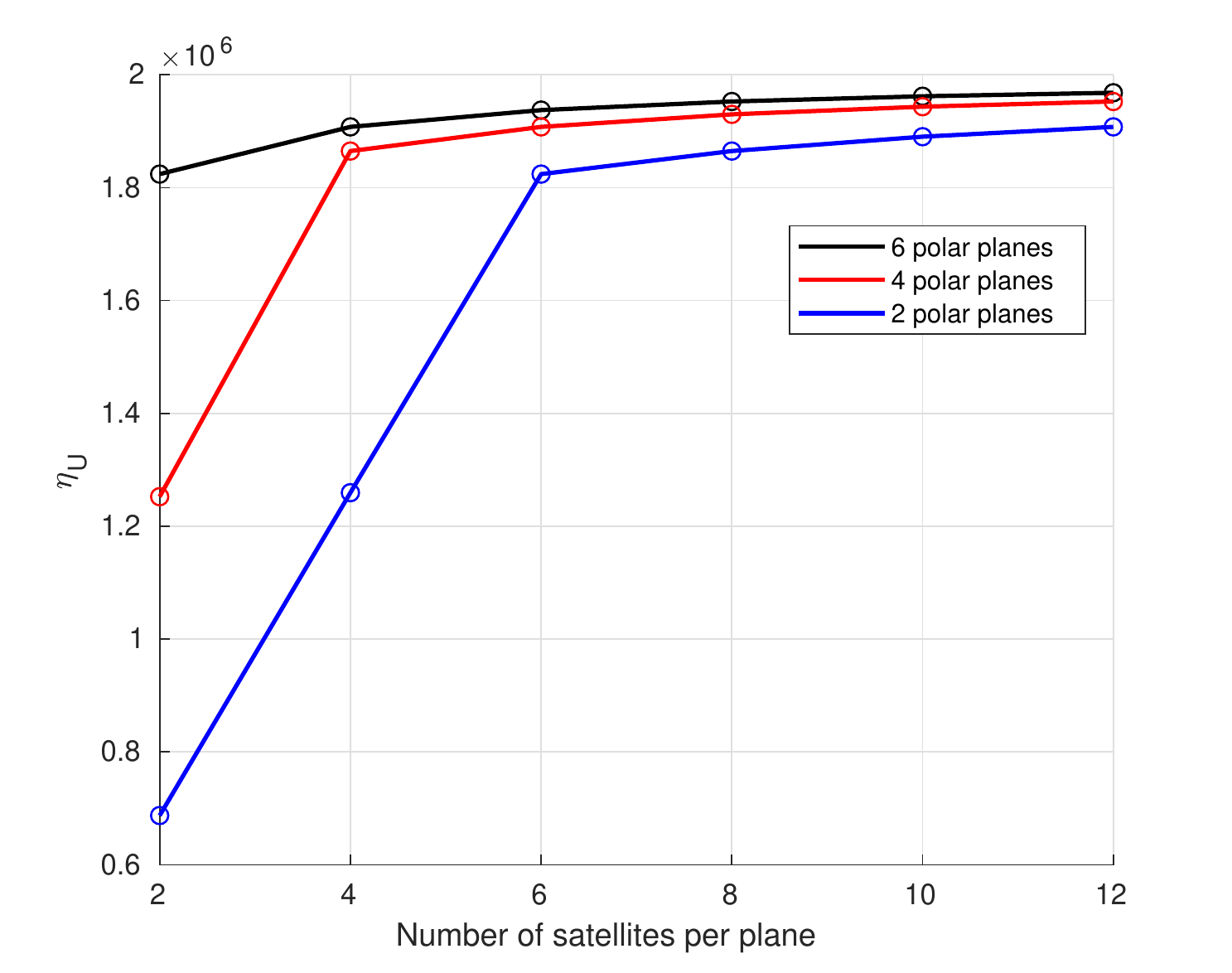}
    \caption{}
    \label{fig:14}
     \end{subfigure}
      \begin{subfigure}[b]{0.475\textwidth}
    \includegraphics[width=3.5in]{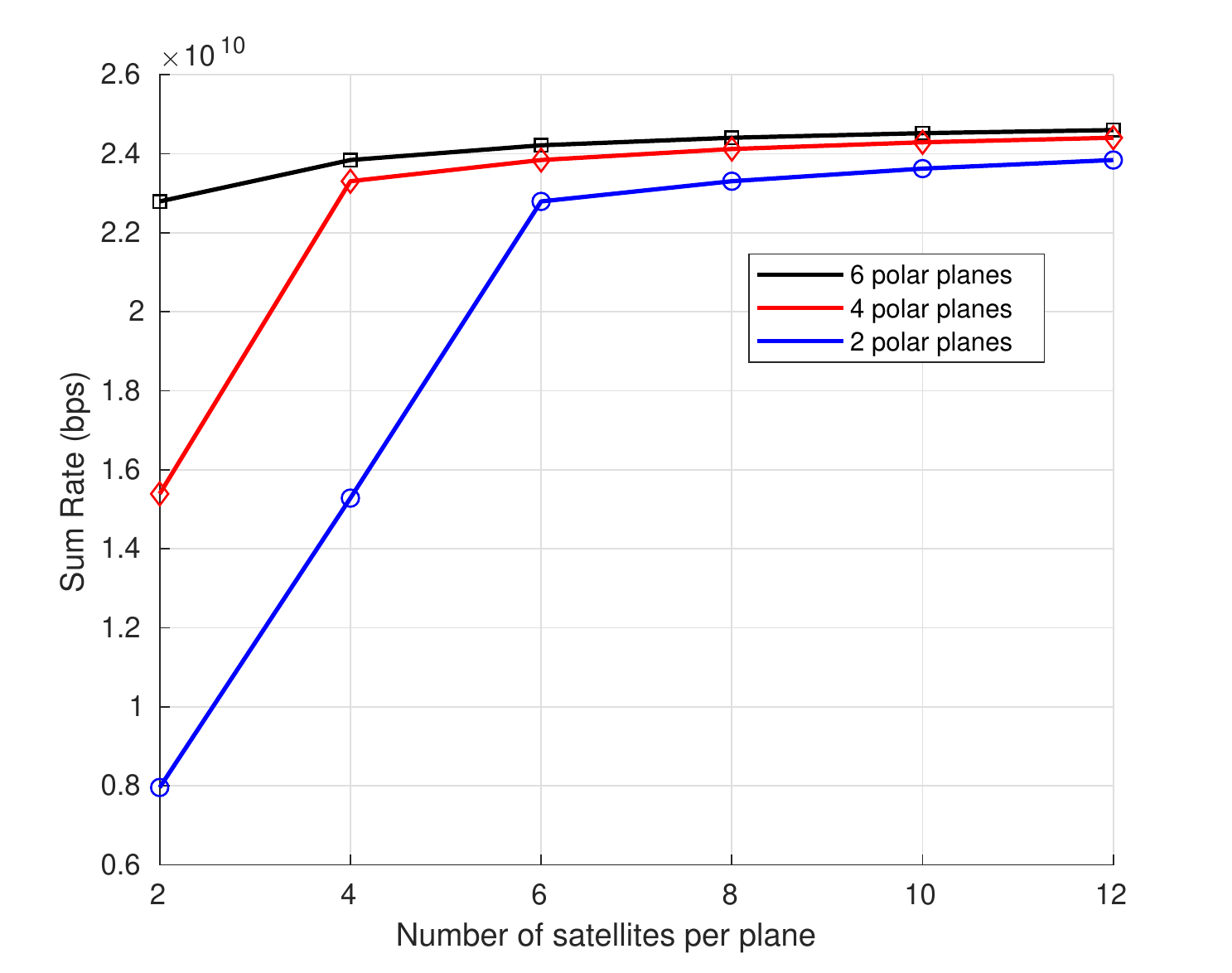}
    \caption{}
    \label{fig:17}
 \end{subfigure}
    \begin{subfigure}[b]{0.475\textwidth}
    \includegraphics[width=3.5in]{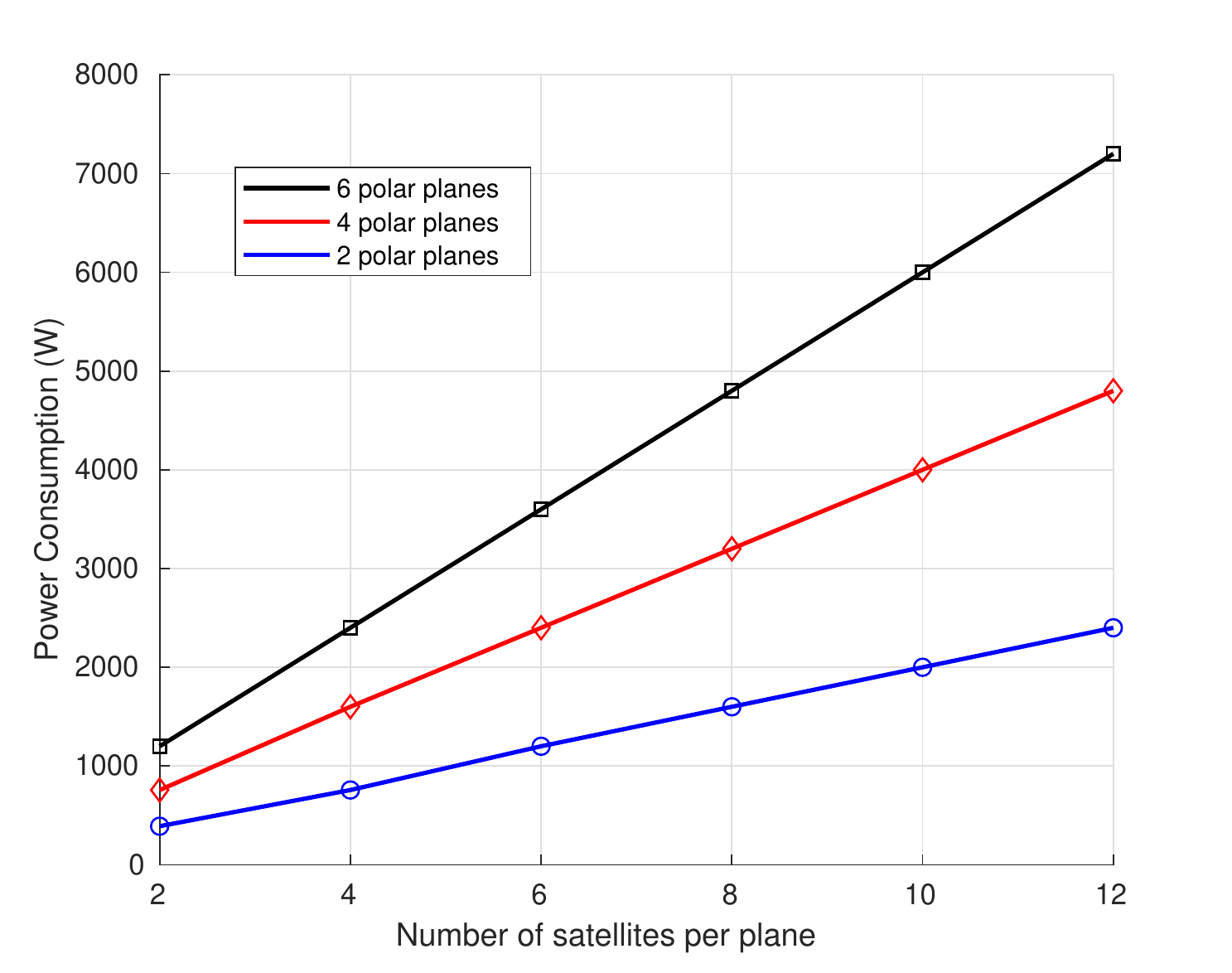}
    \caption{}
    \label{fig:18}
 \end{subfigure}
 \caption{Variation of the Number of Satellites per Polar Plane.}
\end{figure*}

\subsection{ Impact of the Number of Satellites }
In the first simulation, we study the performance of our scheme according to   the number of satellites present per polar plane. We note that we conducted the same simulations for the inclined planes and we obtained analogous results. First, we evaluate the inherent trade-off in offloading between energy  and latency. Therefore, we study jointly the average queue length of the  mMTC offloading queues given by $\sum_{i=1}^I\overline{Q_{i,m}^\text{Sat}}$  and the utility function $\eta_U$ as function of  the number of satellite per polar plane for different  total number of polar planes. 

As depicted in Fig.~\ref{fig:16}, we observe that  the queue length decreases with more satellites in service per polar plane  since more servers are present in the system.  We notice also, for this specific network setting, that placing 2 satellites in 6 polar planes gives the same latency as placing 6 satellites in 2 polar planes. However, we need to place 4 satellites in 4 polar planes to reach the same queue length.   As depicted in Fig.~\ref{fig:14}, we observe that   $\eta_U$ increases   with more satellites present in IMSTN.  These results are supported by the increasing sum rate and the increasing energy consumption presented respectively in Fig.~\ref{fig:17} and Fig.~\ref{fig:18}. 
We note also from Fig.~\ref{fig:18}, that the power consumed by 24 satellites placed as 12 satellites per 2 polar planes is lower compared to 6 satellites per 4 polar planes or 4 satellites per 6 polar planes. The sum rate depicted in Fig.~\ref{fig:18} and the utility function depicted in Fig.~\ref{fig:14} are very close though for the three settings. This observation highlights the importance of the satellites positioning and cooperation to achieve the same performance with less power consumption.

\subsection{Effect of the Trade-off Parameter $\beta$}
In the second simulation, we investigate the influence of the trade-parameter $\beta$ on the performance of our scheme. As defined in equation (\ref{eq:23}), $0\leq \beta \leq 1$ is a  {weighting factor that prioritize the power consumption or the sum rate maximization}. Therefore, we study the mean power consumption and the sum rate variations as a function of $\beta$. 
\begin{figure*}
    \centering
     \begin{subfigure}[b]{0.475\textwidth}
    \centering
    \includegraphics[width=3.5in]{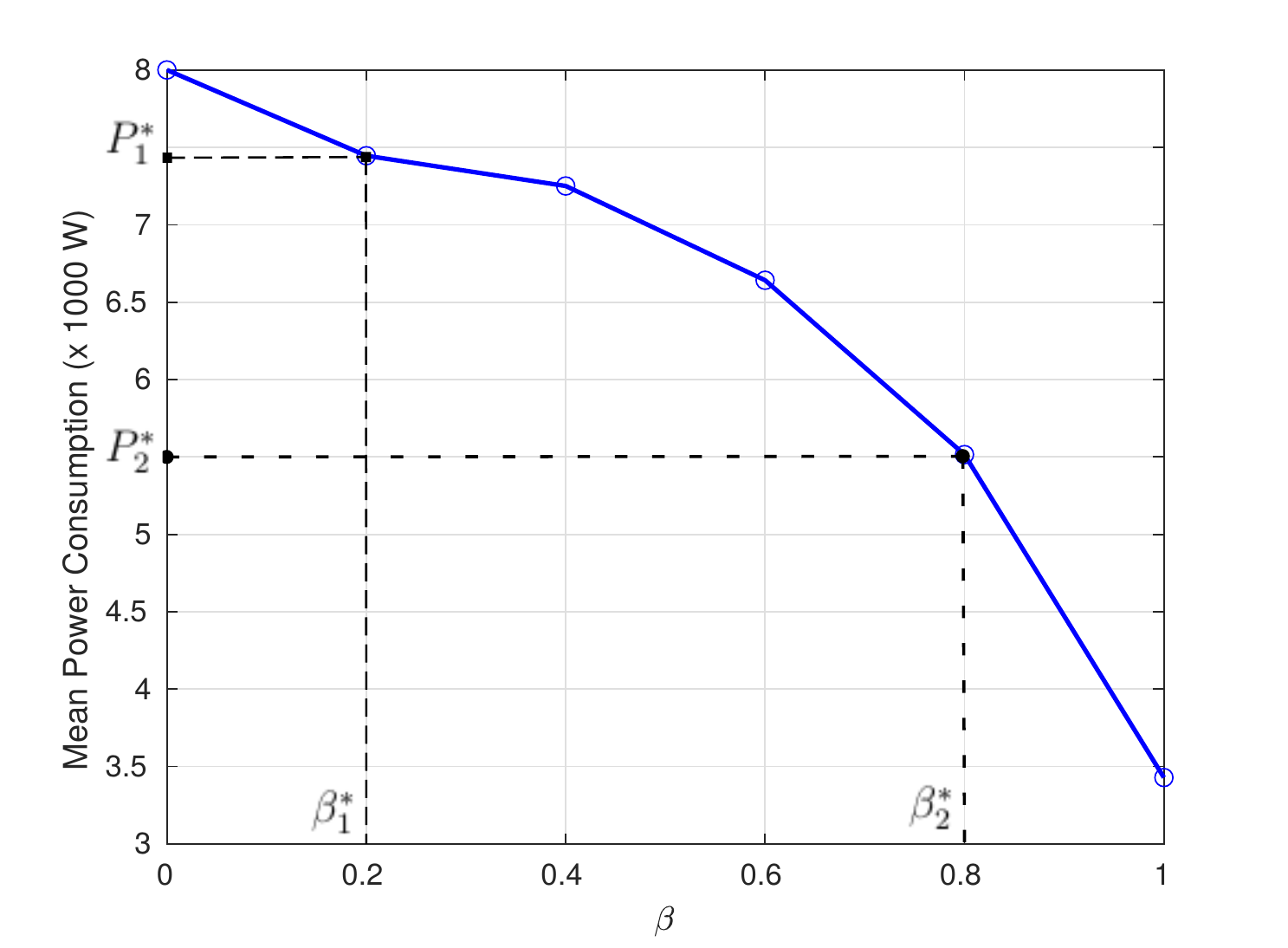}
    \caption{}
    \label{fig:11}
     \end{subfigure}
 \begin{subfigure}[b]{0.475\textwidth}
    \includegraphics[width=3.5in]{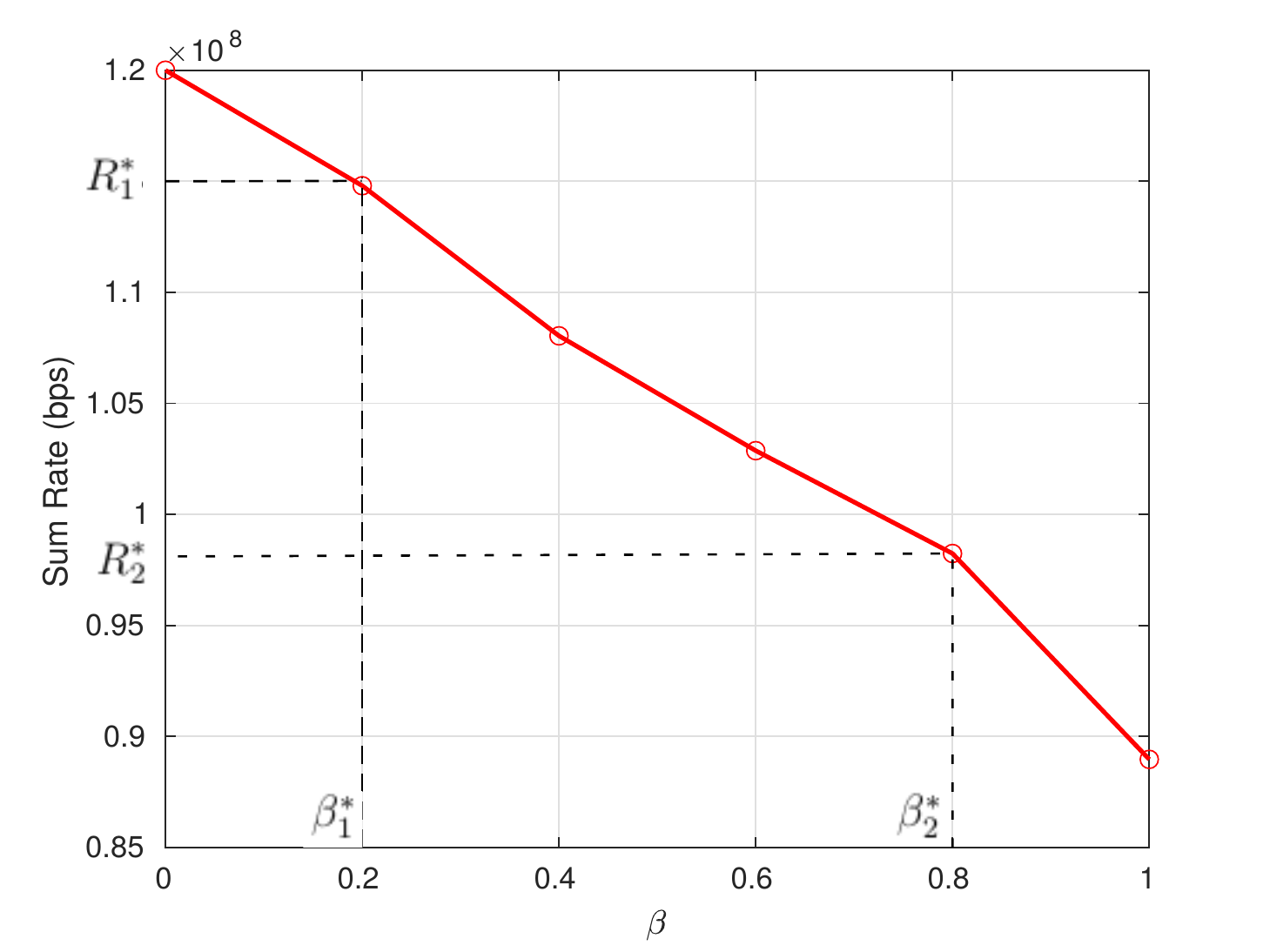}
    \caption{}
    \label{fig:10}
    \end{subfigure}
 \caption{ {Variation of the Trade-off Parameter $\beta$.}}
\end{figure*}
 {When $\beta =0$, the utility function optimization converts to the satellites' power consumption minimization. However,  when $\beta =1$, the optimization problem is equivalent to the satellites' sum rate maximization.} As depicted in Fig.~\ref{fig:11}, we observe that the more $\beta$ increases, the more the consumed power decreases until reaching its minimum for $\beta=1$. This observation is supported by (\ref{eq:23}). As depicted in Fig.~\ref{fig:10}, we observe that the more $\beta$ increases, the more the sum rate decreases. This observation is due to the fact that less power is allocated for higher values of $\beta$ as suggested by Fig.~\ref{fig:11}.
 {Based on these findings, $\beta$ can be fixed practically through two options. First, when the maximum consumed power of the satellites is fixed for example at $\sum_{j=1}^M  \sum_{i=1}^I \gamma_j P_{ij}(t)=P_1^*$.  Then, we set $\beta=\beta_1^*$ based on the corresponding power $P_1^*$ from Fig.~6(a). We can assess the achievable satellites' throughput accordingly from Fig.~6(b) $R^*_1$. Second, when the desired throughput that the satellites should achieve is fixed for example at $\sum_{i=1}^I\sum_{j=1}^M R_{ij}(t)=R_2^*$. Then, we set $\beta=\beta_2^*$ based on the corresponding sum rate $R_2^*$ from Fig.~6(b). We can assess the total consumed power $P_2^*$ accordingly from   Fig.~6(a).}
\subsection{Effect of the Trade-off Parameter $V$}
\begin{figure*}[b!]
    \centering
     \begin{subfigure}[b]{0.473\textwidth}
   \includegraphics[width=3.5in]{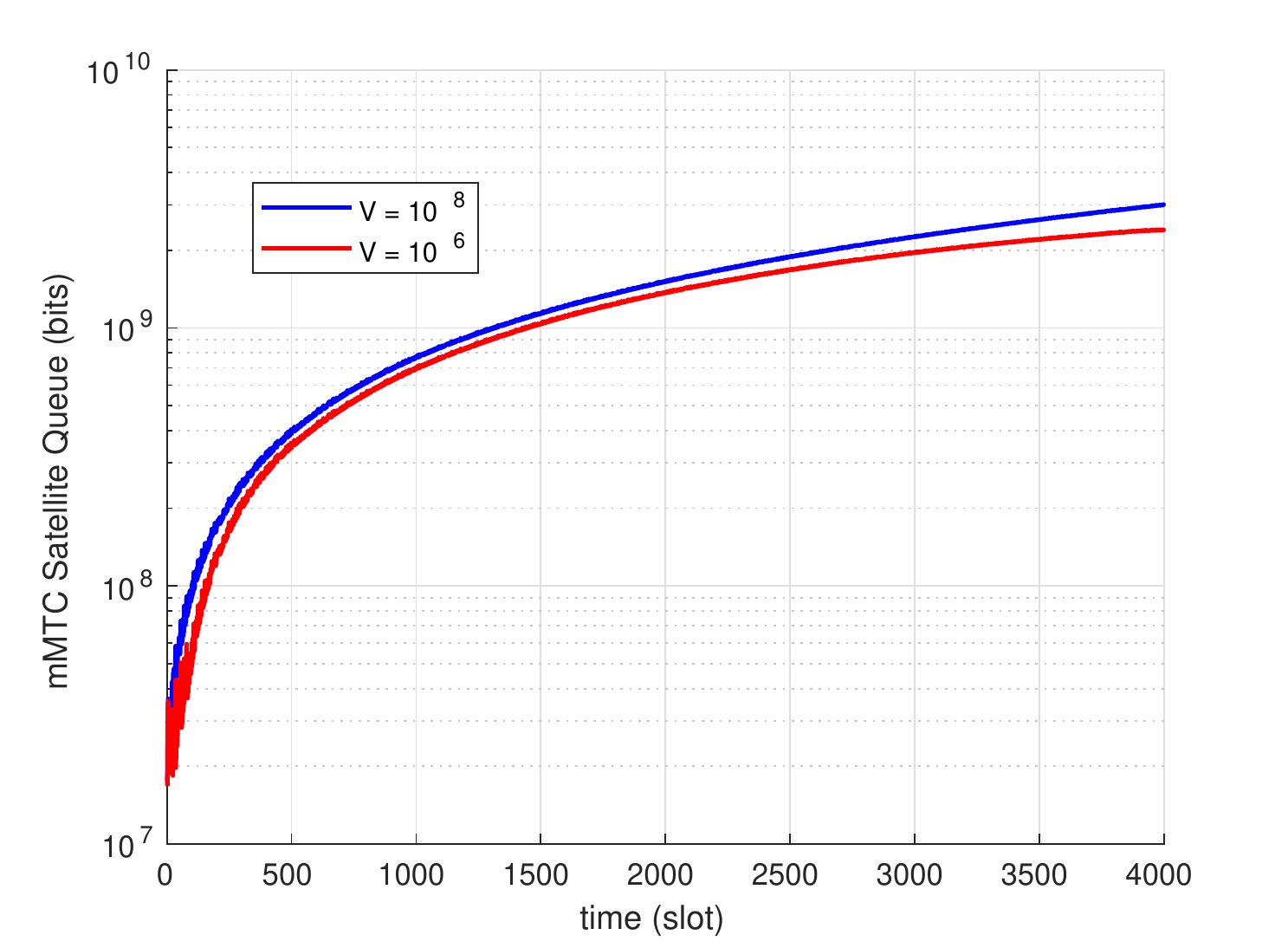}
    \caption{}
    \label{fig:7}
     \end{subfigure}
           \begin{subfigure}[b]{0.475\textwidth}
\includegraphics[width=3.5in]{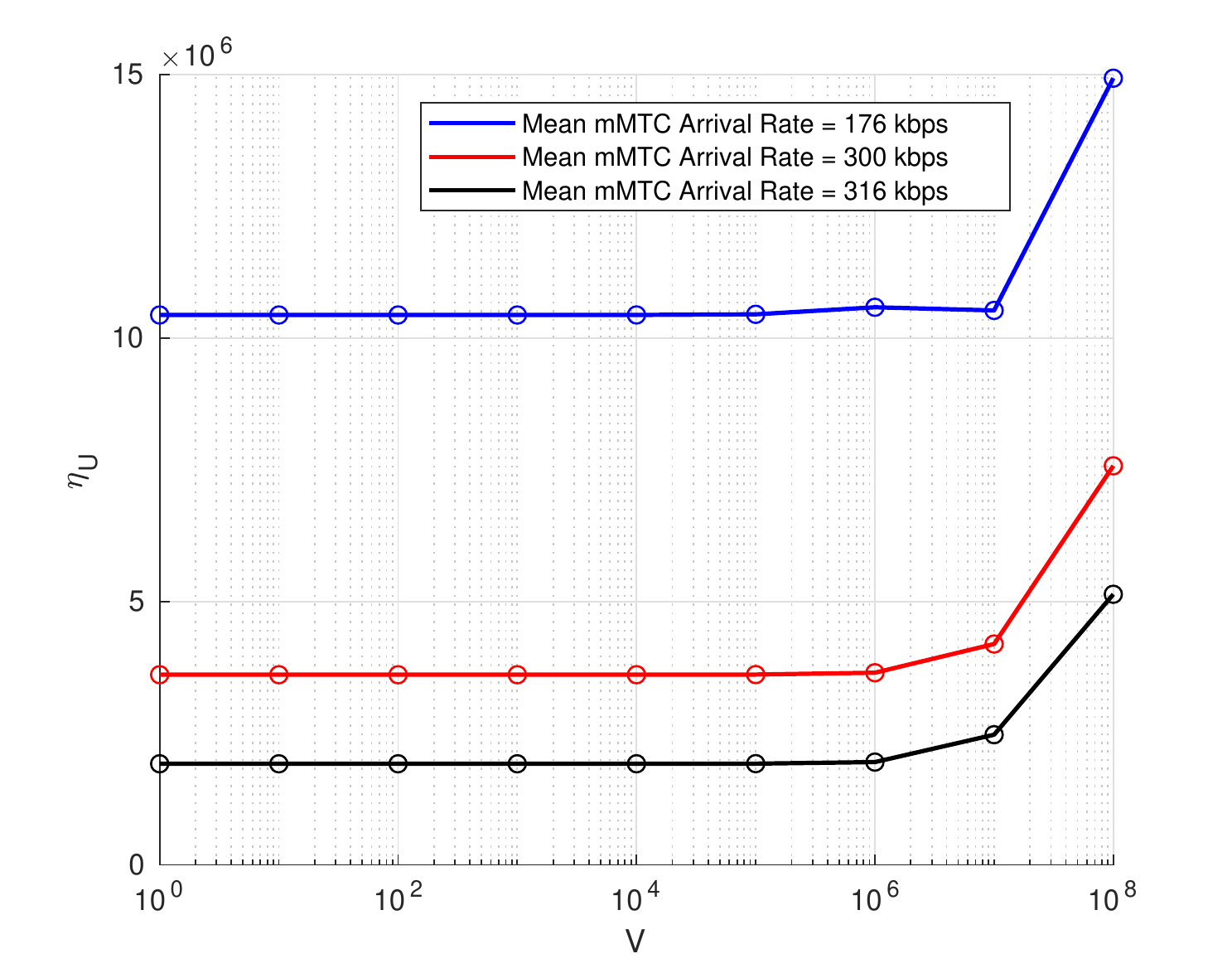}
    \caption{}
    \label{fig:6}
 \end{subfigure}
     \begin{subfigure}[b]{0.475\textwidth}
    \includegraphics[width=3.5in]{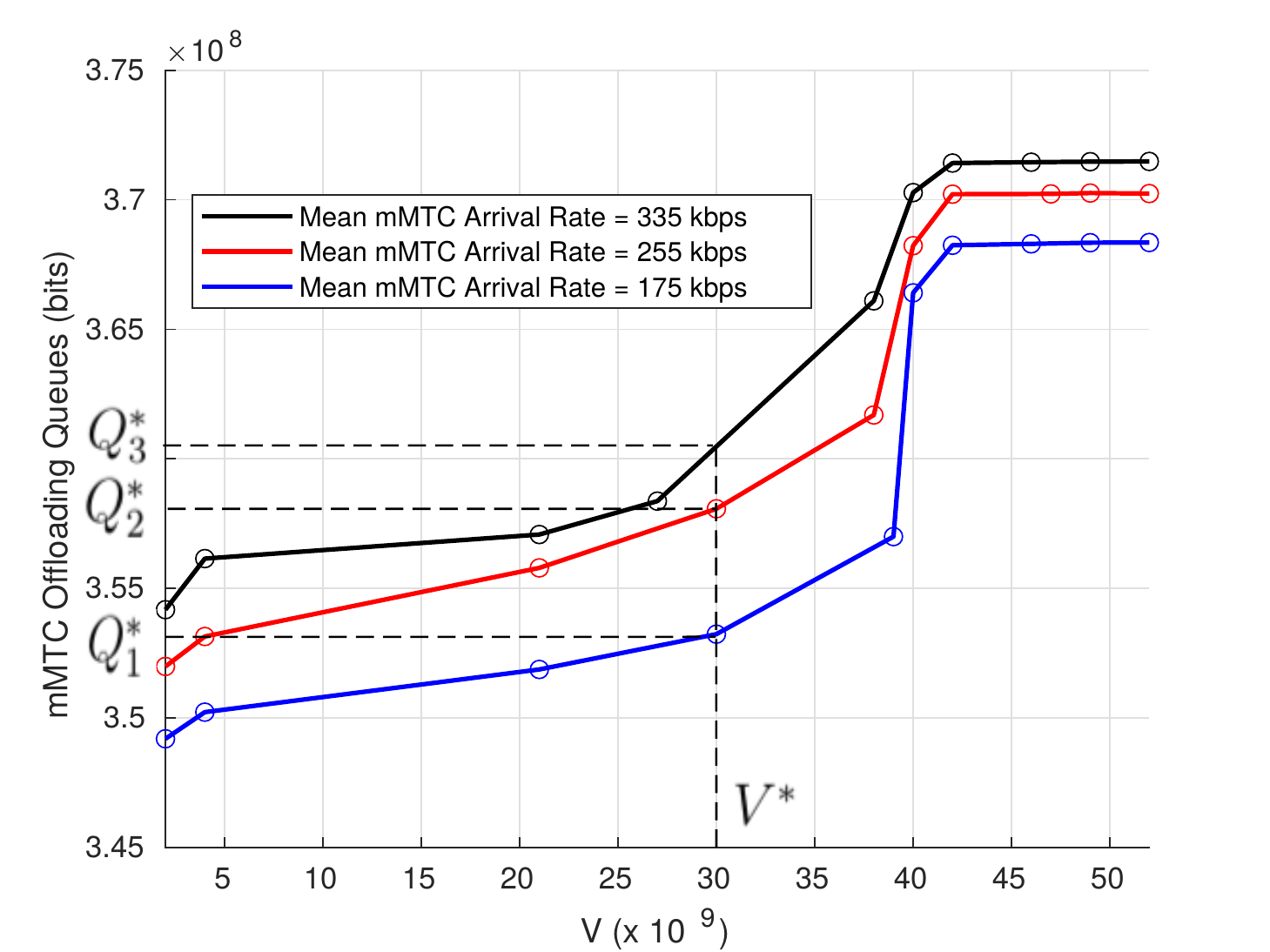}
    \caption{}
    \label{fig:8}
 \end{subfigure}
    \begin{subfigure}[b]{0.475\textwidth}
   \includegraphics[width=3.5in]{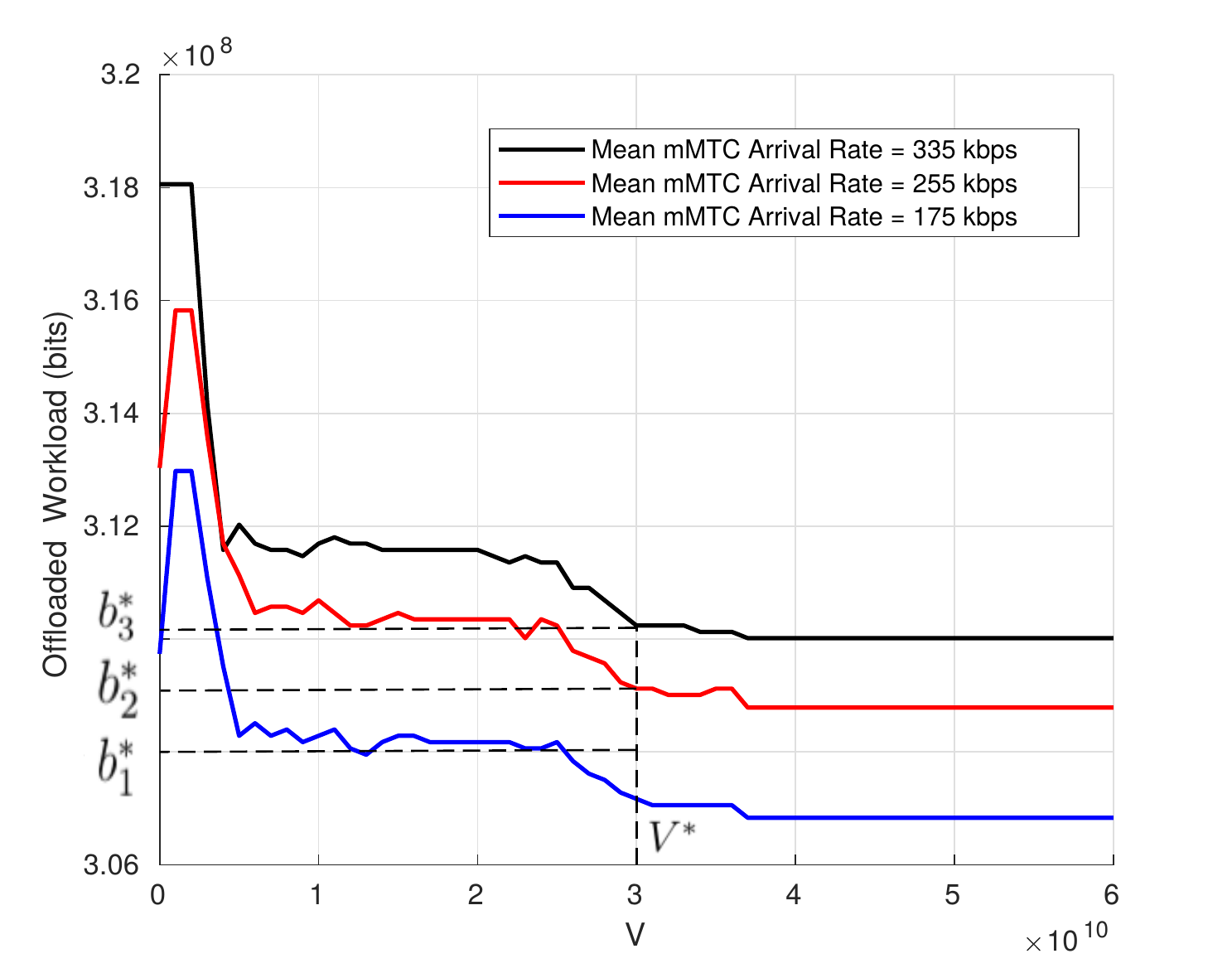}
    \caption{}
    \label{fig:9}
 \end{subfigure}
 \caption{ {Variation of the trade-off parameter $V$.}}
\end{figure*}

In the third simulation, we study the performance of our scheme according to  the  trade-off parameter $V$, which is defined in the drift-plus-penalty (c.f. (\ref{eq:drift})) and configures how much we emphasize on $\eta_U$ maximization versus the latency. 


First, we evaluate the queue stability of our offloading scheme. Therefore, we study the mMTC satellite average queue length given by $\sum_{i=1}^I \overline{Q_{i,m}^\text{Sat}}$ , as function of $t$ for two values of $V$. As depicted in Fig.~\ref{fig:7}, we observe that the queue length increases with $t$ at the beginning until reaching a plateau for larger values of $t$. We also notice that a larger value of $V$ leads as expected to larger stable values of  the queue length \cite{neely2010stochastic}.

Secondly, we evaluate the inherent trade-off in offloading between energy  and latency. Therefore, we study {jointly}  the utility function $\eta_U$ and the average queue length of the mMTC offloading queues   as function of $V$ for different values of mMTC arrival rate. 
First, we examine the utility function $\eta_U$ for a wide range of $V$  varying from $1$ to $10^8$. As depicted in Fig.~\ref{fig:6}, we observe that the more  $V$ increases, the more $\eta_U$ increases  for the different arrival rates because we emphasize  on the utility function. We notice also that, for the same value of $V$,   $\eta_U$ decreases with higher arrival rates. This observation is due to the fact that more traffic is accepted into the network with higher arrival rates. Afterwards, we investigate the experienced latency during offloading by assessing the length of the mMTC offloading queues given by $\sum_{i=1}^I\overline{\big(Q_{i,m}^\text{Ter}+Q_{i,m}^\text{Sat}\big)}$. We study the latency  for a narrower range of $V$ based on the result of the utility function depicted in Fig.~\ref{fig:6}. 
As depicted in Fig.~\ref{fig:8}, we observe that the more  $V$ increases, the more the queue length increases for the different arrival rates as anticipated \cite{neely2010stochastic}.  We notice also that, for the same value of $V$, the queue length increases with higher arrival rates. This observation is due to the fact that more traffic is accepted into the network with higher arrival rates.  Consequently, the offloading scheme should sacrifice the utility function  to keep the queues stable as supported by Fig.~\ref{fig:6}.

Then, we examine the offloading decisions of our scheme. Therefore, we study the average mMTC offloaded workload given by $\sum_{i=1}^I \overline{\big(b_{i,m}^\text{Ter}+b_{i,m}^\text{Sat}\big)}$ as function of $V$ for different values of mMTC arrival rate. As depicted in Fig.~\ref{fig:9}, we observe that the offloaded workload reaches its maximum for $V$ around $2.5\times 10^9$. This observation  {is due to the fact that the offloading} decisions depend not only on $V$, but also on the queues length as supported by (\ref{eq:47'}) and (\ref{eq:48'}). We observe that more traffic is offloaded to both backhauls with higher mMTC arrival rates, which proves the ability of our scheme to cover more users and offer better network availability.  

 {Practically, $V$ can be selected under the condition of stable queues by verifying that the queue length reaches the stable regime over time as depicted in Fig.~7(a). Then, a specific value of $V^*$ can be fixed based on a given tolerable delay assessed through the queue length $Q^*$  or  a desired offloaded workload $b^*$ for a fixed arrival rate. $V^*$ can be determined based on Fig.~7(c) and Fig.~7(d) respectively for a fixed delay and a fixed offloaded workload.}

 \begin{figure*}[b!]
    \centering
     \begin{subfigure}[b]{0.475\textwidth}
    \includegraphics[width=3.5in]{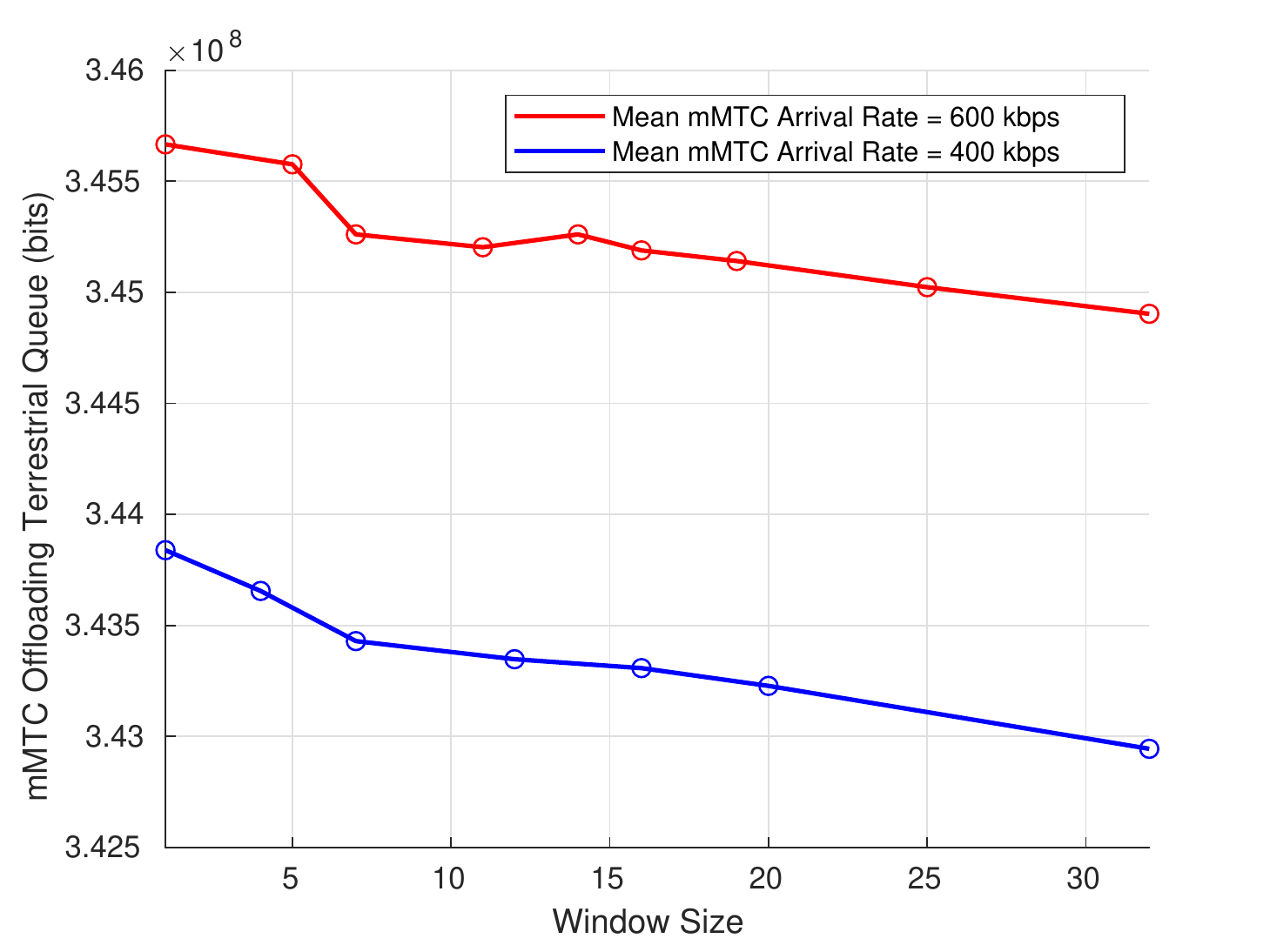}
    \caption{}
    \label{fig:25}
     \end{subfigure}
    \begin{subfigure}[b]{0.475\textwidth}
\includegraphics[width=3.5in]{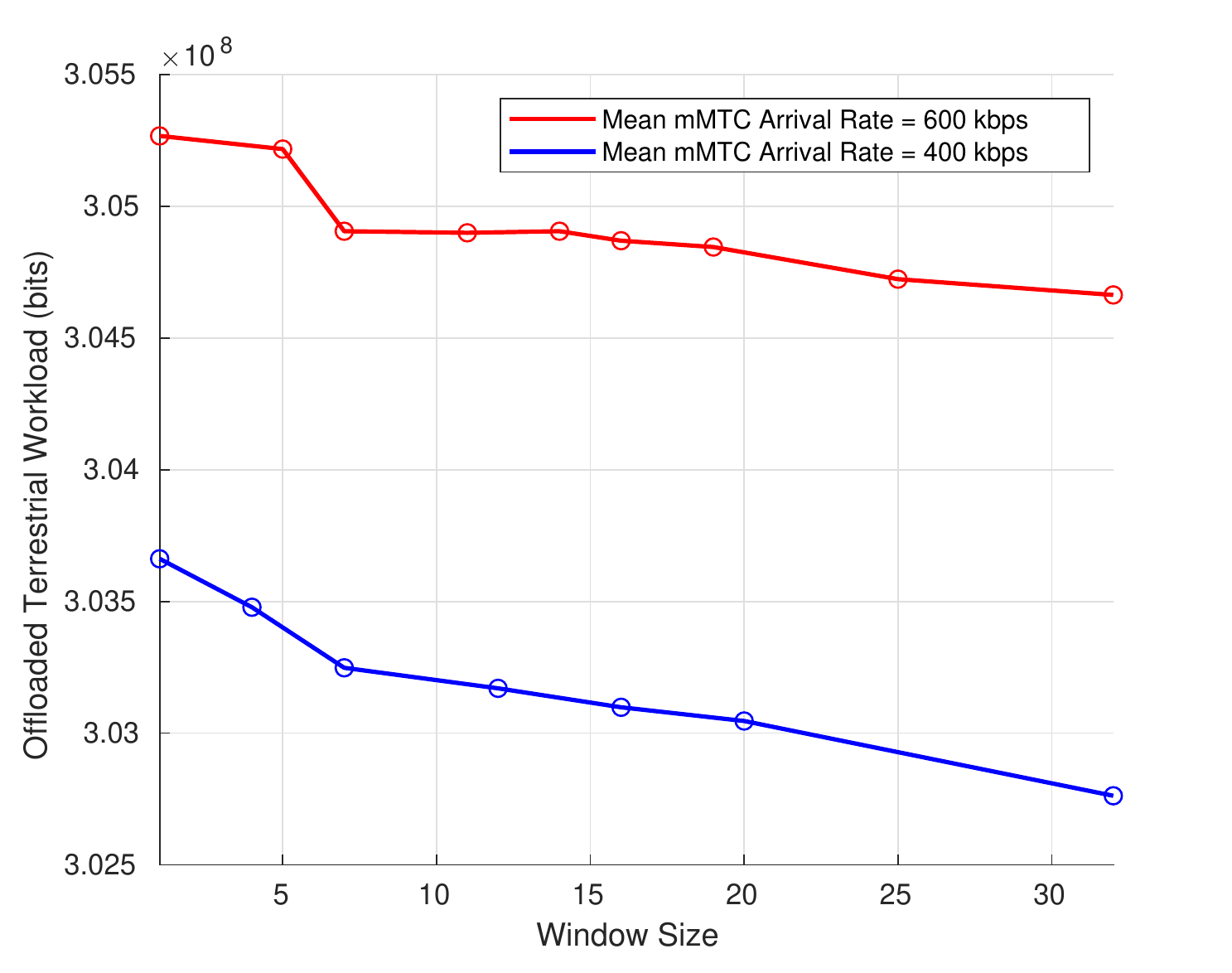}
    \caption{}
    \label{fig:24}
 \end{subfigure}
 \caption{Variation of the Window Size.}
\end{figure*}
\subsection{ Variation of the Window Size }
In the last simulation, we study the performance of our scheme according to the traffic prediction window size. 

We study the mMTC terrestrial average queue length given by $\sum_{i=1}^I \overline{Q_{i,m}^\text{Ter}}$, as function of the window size for two values of mMTC arrival rate. As depicted in Fig.~\ref{fig:25}, we observe that the queue length decreases with  larger window sizes thanks to traffic prediction. This observation highlights the importance of traffic prediction to relieve the congestion in the terrestrial backhaul, since future arrival can be pre-served.  We also notice that a higher arrival rate leads as expected to  a longer queue and hence to higher latency. Moreover, this improvement comes at no cost in terms of power consumption since our scheme achieves a constant total power   and a constant $\eta_U$ for different window sizes based on the simulations we conducted. However, less mMTC traffic can be offloaded to the terrestrial backhaul, as supported by Fig.~\ref{fig:24}, to preserve the queues stability and the reliability and latency requirements.

\section{Conclusion}
In this paper, we proposed a proactive offloading scheme in dynamic IMSTN. Our proposed offloading scheme optimizes and  steers the eMBB traffic to the satellite backhaul, the  URLLC  traffic  to  the  terrestrial  backhaul  and  the mMTC  traffic  to  both  backhauls  while  answering  their heterogeneous requirements. Our  offloading  scheme has  recourse to traffic prediction to pre-allocate the necessary resources and answer the different requirements of these heterogeneous traffic types present in IMSTN in terms of data-rate, latency and reliability. 
 Our  scheme succeeds jointly to meet these requirements and to reduce the consumed energy during offloading  by establishing and balancing first a trade-off between energy consumption and the achievable data-rate and second a trade-off between energy consumption and latency. Our findings highlight the pivotal role that traffic prediction plays {to balance these fundamental trade-offs} and to relieve the congestion in the terrestrial backhaul while optimizing the power~consumption.

 \section*{Appendix}
 Based on Lemma 4.6 of the Min Drift-Plus-Penalty Algorithm \cite{neely2010stochastic}, the drift-plus-penalty has the following upper bound ${\forall\;}t, \;{\forall\;}V>0, \;{\forall\;}\Theta(t)$:
 \begin{footnotesize}
\begin{align}
 &\Delta(\Theta(t))-V\mathop{\mathbb{E}}[\eta_U(t)|\Theta(t)]\leq B-V\mathop{\mathbb{E}}[\eta_U(t)|\Theta(t)] 
    +\sum_{i=1}^I{Q_{i,m}^{sum}(t)\mathop{\mathbb{E}}\big[A_{i,m}(t+W_i)-\big(b_{i,m}^\text{Ter}(t)+b_{i,m}^\text{Sat}(t)\big)\big|\Theta(t)\big]}\nonumber\\
    &+ \sum_{i=1}^I{Q_{i,m}^\text{Ter}(t)\mathop{\mathbb{E}}\big[b_{i,m}^\text{Ter}(t)-\big(C_i^\text{Ter}(t)-A_{i,u}(t)\big)\big|\Theta(t)\big]}
    + \sum_{i=1}^I{Q_{i,m}^\text{Sat}(t)\mathop{\mathbb{E}}\big[b_{i,m}^\text{Sat}(t)-\big(\sum_{j=1}^M R_{ij}(t)-A_{i,e}(t)\big)\big|\Theta(t)\big]}\nonumber\\
     &+ \sum_{i=1}^I{Q_{i,u}^\text{Ter}(t)\mathop{\mathbb{E}}\big[A_{i,u}(t)-\big(C_i^\text{Ter}(t)-b_{i,m}^\text{Ter}(t)\big)\big|\Theta(t)\big]}
     + \sum_{i=1}^I{Q_{i,e}^\text{Sat}(t)\mathop{\mathbb{E}}\big[A_{i,e}(t)-\big(\sum_{j=1}^M R_{ij}(t)-b_{i,m}^\text{Sat}(t)\big)\big|\Theta(t)\big]}\nonumber\\
    &+ \sum_{i=1}^I{Z_{i,m}^\text{Ter}(t)\mathop{\mathbb{E}}\big[\mathds{1}_{\{Q_{i,m}^\text{Ter}(t+1)> D_i^{m,\nu}\}}-\epsilon_{m}^\text{Ter}\big|\Theta(t)\big]}
    + \sum_{i=1}^I{Z_{i,m}^\text{Sat}(t)\mathop{\mathbb{E}}\big[\mathds{1}_{\{Q_{i,m}^\text{Sat}(t+1)> D_i^{m,\nu}\}}-\epsilon_{m}^\text{Sat}\big|\Theta(t)\big]}\nonumber\\
    &+ \sum_{i=1}^I{Z_{i,u}^\text{Ter}(t)\mathop{\mathbb{E}}\big[\mathds{1}_{\{Q_{i,u}^\text{Ter}(t+1)> D_i^u\}}-\epsilon_{u}\big|\Theta(t)\big]}
 \end{align}
 \end{footnotesize}
 where $B$ is a positive constant that satisfies the following for all $t$:
 
  \begin{footnotesize}
 \begin{align}
     B\geq& \sum_{i=1}^I{\mathop{\mathbb{E}}\big[A_{i,m}(t)^2+\big(b_{i,m}^\text{Ter}(t)+b_{i,m}^\text{Sat}(t)\big)^2\big|\Theta(t)\big]}+ \sum_{i=1}^I{\mathop{\mathbb{E}}\big[b_{i,m}^\text{Ter}(t)^2+\big(C_i^\text{Ter}(t)-A_{i,u}(t)\big)^2\big|\Theta(t)\big]}\nonumber\\
    &+ \sum_{i=1}^I{\mathop{\mathbb{E}}\big[b_{i,m}^\text{Sat}(t)^2+\big(\sum_{j=1}^M R_{ij}(t)-A_{i,e}(t)\big)^2\big|\Theta(t)\big]}+ \sum_{i=1}^I{\mathop{\mathbb{E}}\big[A_{i,u}(t)^2-\big(C_i^\text{Ter}(t)-b_{i,m}^\text{Ter}(t)\big)^2\big|\Theta(t)\big]}\nonumber\\
     &+ \sum_{i=1}^I{\mathop{\mathbb{E}}\big[A_{i,e}(t)^2-\big(\sum_{j=1}^M R_{ij}(t)-b_{i,m}^\text{Sat}(t)\big)^2\big|\Theta(t)\big]}+ \sum_{i=1}^I{\mathop{\mathbb{E}}\big[\big(\mathds{1}_{\{Q_{i,m}^\text{Ter}(t+1)> D_i^{m,\nu}\}}-\epsilon_{m}^\text{Ter}\big)^2\big|\Theta(t)\big]}\nonumber\\
    &+ \sum_{i=1}^I{\mathop{\mathbb{E}}\big[\big(\mathds{1}_{\{Q_{i,m}^\text{Sat}(t+1)> D_i^{m,\nu}\}}-\epsilon_{m}^\text{Sat}\big)^2\big|\Theta(t)\big]}+ \sum_{i=1}^I{\mathop{\mathbb{E}}\big[\big(\mathds{1}_{\{Q_{i,u}^\text{Ter}(t+1)> D_i^u\}}-\epsilon_{u}\big)^2\big|\Theta(t)\big]}
 \end{align}
  \end{footnotesize}

 Since $\mathop{\mathbb{E}}[A_{i,\kappa}(t)]=\lambda_i^\kappa$ and by using (\ref{eq:23}), the upper bound of the drift-plus-penalty can be simplified as follows:
 
   \begin{footnotesize}
 \begin{align}
 \Delta(\Theta&(t))-V\mathop{\mathbb{E}}[\eta_U(t)|\Theta(t)]\leq B -\frac{V\beta}{I\times M}\sum_{j=1}^M\sum_{i=1}^I \sigma_i \mathop{\mathbb{E}}[R_{ij}(t)|\Theta(t)]+\frac{V(1-\beta)}{I \times M}\sum_{j=1}^M  \sum_{i=1}^I\gamma_j\mathop{\mathbb{E}}[ P_{ij}(t)|\Theta(t)] \nonumber \\
    &+\sum_{i=1}^I{Q_{i,m}^\text{sum}(t)\bigg(\lambda_i^m-\mathop{\mathbb{E}}\big[b_{i,m}^\text{Ter}(t)+b_{i,m}^\text{Sat}(t)\big|\Theta(t)\big]\bigg)}+ \sum_{i=1}^I{Q_{i,m}^\text{Ter}(t)\bigg(\lambda_i^u+\mathop{\mathbb{E}}\big[b_{i,m}^\text{Ter}(t)-C_i^\text{Ter}(t)\big|\Theta(t)\big]\bigg)}\nonumber\\
    &+ \sum_{i=1}^I{Q_{i,m}^\text{Sat}(t)\bigg(\lambda_i^e+\mathop{\mathbb{E}}\big[b_{i,m}^\text{Sat}(t)-\sum_{j=1}^M R_{ij}(t)\big|\Theta(t)\big]\bigg)}+ \sum_{i=1}^I{Q_{i,u}^\text{Ter}(t)\bigg(\lambda_i^u+\mathop{\mathbb{E}}\big[b_{i,m}^\text{Ter}-C_i^\text{Ter}(t)\big|\Theta(t)\big]\bigg)}\nonumber\\
     &+ \sum_{i=1}^I{Q_{i,e}^\text{Sat}(t)\bigg(\lambda_i^e+\mathop{\mathbb{E}}\big[b_{i,m}^\text{Sat}(t)-\sum_{j=1}^M R_{ij}(t)\big|\Theta(t)\big]\bigg)}+ \sum_{i=1}^I{Z_{i,m}^\text{Ter}(t)\mathop{\mathbb{E}}\big[\mathds{1}_{\{Q_{i,m}^\text{Ter}(t+1)> D_i^{m,\nu}\}}-\epsilon_{m}^\text{Ter}\big|\Theta(t)\big]}\nonumber\\
    &+ \sum_{i=1}^I{Z_{i,m}^\text{Sat}(t)\mathop{\mathbb{E}}\big[\mathds{1}_{\{Q_{i,m}^\text{Sat}(t+1)> D_i^{m,\nu}\}}-\epsilon_{m}^\text{Sat}\big|\Theta(t)\big]}+ \sum_{i=1}^I{Z_{i,u}^\text{Ter}(t)\mathop{\mathbb{E}}\big[\mathds{1}_{\{Q_{i,u}^\text{Ter}(t+1)> D_i^u\}}-\epsilon_{u}\big|\Theta(t)\big]}
    \label{eq:48}
 \end{align}
   \end{footnotesize}

 Instead of minimizing the drift-plus-penalty, we seek to minimize the upper bound (\text{UB}) expanded in (\ref{eq:48}). However, it is hard to solve an optimization problem with expectation. Therefore, we approximately solve our problem (\ref{eq:39''}) by adopting the framework of opportunistically minimizing a conditional expectation \cite{neely2010stochastic,gao2019pora,peng2016energy}. Accordingly, our problem (\ref{eq:39''}) is approximated to the following problem: 
 
    \begin{footnotesize}
\begin{align}
\label{eq:480}
\min_{b,\vect{P}}  \quad 
&\frac{V(1-\beta)}{I\times M}\sum_{j=1}^M  \sum_{i=1}^I \gamma_j \frac{P_{ij}(t)}{P_j^\text{max}}-\frac{V\beta}{I \times M}\sum_{j=1}^M\sum_{i=1}^I \sigma_i \frac{R_{ij}(t)}{C^\text{Sat}}+\sum_{i=1}^Ib_{i,m}^\text{Ter}(t) \bigg(Q_{i,m}^\text{Ter}(t)+Q_{i,u}^\text{Ter}(t)-Q_{i,m}^\text{Sum}(t)\bigg)\\
&+\sum_{i=1}^Ib_{i,m}^\text{Sat}(t) \bigg(Q_{i,m}^\text{Sat}(t)+Q_{i,e}^\text{Sat}(t)-Q_{i,m}^\text{Sum}(t)\bigg)-\sum_{i=1}^I\sum_{j=1}^M R_{ij}(t)\bigg(Q_{i,m}^\text{Sat}(t)+Q_{i,e}^\text{Sat}(t)\bigg)\nonumber \\
&+\sum_{i=1}^I Z_{i,m}^\text{Ter}(t)\;\mathds{1}_{\big\{\big[Q_{i,m}^\text{Ter}(t)-C_i^\text{Ter}(t)+A_{i,u}(t)\big]^++b_{i,m}^\text{Ter}(t)> D_i^{m,\nu}\big\}}\nonumber \\
&+\sum_{i=1}^IZ_{i,m}^\text{Sat}(t)\;\mathds{1}_{\big\{\big[Q_{i,m}^\text{Sat}(t)-\sum_{j=1}^M R_{ij}(t)+A_{i,e}(t)\big]^++b_{i,m}^\text{Sat}(t)> D_i^{m,\nu}\big\}}\nonumber \\
&+\sum_{i=1}^IZ_{i,u}^\text{Ter}(t)\;\mathds{1}_{\big\{\big[Q_{i,u}^\text{Ter}(t)-C_i^\text{Ter}(t)+b_{i,m}^\text{Ter}(t)\big]^++A_{i,u}(t)> D_i^u\big\}}   \nonumber \\
&\mathrm{s.t.} \quad C1-C4 \nonumber 
\end{align}
     \end{footnotesize}
Using the approximation in \cite{liu2019dynamic}, we equivalently re-write the minimization problems as:
    \begin{footnotesize}
\begin{align}
\footnotesize
  &\min_{b_{i,m}^\text{Ter}}   Z_{i,m}^\text{Ter}(t)\;\mathds{1}_{\big\{\big[Q_{i,m}^\text{Ter}(t)-C_i^\text{Ter}(t)+A_{i,u}(t)\big]^++b_{i,m}^\text{Ter}(t)> D_i^{m,\nu}\big\}}  \equiv  \min_{b_{i,m}^\text{Ter}}   b_{i,m}^\text{Ter}(t) \bigg(Z_{i,m}^\text{Ter}(t)+Q_{i,m}^\text{Ter}(t)+ A_{i,u}(t)-C_i^\text{Ter}(t)\bigg) 
\end{align} 
\begin{align}
\footnotesize
 &\min_{b_{i,m}^\text{Ter}}  Z_{i,u}^\text{Ter}(t)\;\mathds{1}_{\big\{\big[Q_{i,u}^\text{Ter}(t)-C_i^\text{Ter}(t)+b_{i,m}^\text{Ter}(t)\big]^++A_{i,u}(t)> D_i^u\big\}}  \equiv  \min_{b_{i,m}^\text{Ter}}   b_{i,m}^\text{Ter}(t)\bigg(Z_{i,u}^\text{Ter}(t)+Q_{i,u}^\text{Ter}(t)+ A_{i,u}(t)-C_i^\text{Ter}(t)\bigg)
  \end{align}
\begin{align}
\footnotesize
    &\min_{b_{i,m}^\text{Sat},P_{ij},{B}_{ij}}   \sum_{i=1}^IZ_{i,m}^\text{Sat}(t)\;\mathds{1}_{\big\{\big[Q_{i,m}^\text{Sat}(t)-\sum_{j=1}^M R_{ij}(t)+A_{i,e}(t)\big]^++b_{i,m}^\text{Sat}(t)> D_i^{m,\nu}\big\}} \equiv \nonumber \\ &\min_{b_{i,m}^\text{Sat},P_{ij},{B}_{ij}}   \left(b_{i,m}^\text{Sat}(t)-\sum_{j=1}^M R_{ij}(t)\right)\left(Z_{i,m}^\text{Sat}(t)+Q_{i,m}^\text{Sat}(t)+ A_{i,e}(t)\right) \\
\end{align}  

     \end{footnotesize}

  Rearranging in (\ref{eq:480}), our optimization problem is equivalently re-expressed as:
  
      \begin{footnotesize}
  \begin{align}
  \label{eq:52}
\min_{b,\vect{P}}  \quad &
\frac{V(1-\beta)}{I\times M}\sum_{j=1}^M  \sum_{i=1}^I \gamma_j \frac{P_{ij}(t)}{P_j^\text{max}}-\frac{V\beta}{I \times M}\sum_{j=1}^M
\sum_{i=1}^I \sigma_i \frac{R_{ij}(t)}{C^\text{Sat}} \\
&+\sum_{i=1}^Ib_{i,m}^\text{Ter}(t) \bigg(2Q_{i,m}^\text{Ter}(t)+2Q_{i,u}^\text{Ter}(t)+Z_{i,m}^\text{Ter}(t)+Z_{i,u}^\text{Ter}(t)+ 2A_{i,u}(t) -2C_i^\text{Ter}(t)-Q_{i,m}^\text{Sum}(t)\bigg)\nonumber\\
&+\sum_{i=1}^Ib_{i,m}^\text{Sat}(t) \bigg(2Q_{i,m}^\text{Sat}(t)+Q_{i,e}^\text{Sat}(t)+Z_{i,m}^\text{Sat}(t))+ A_{i,e}(t)-Q_{i,m}^\text{Sum}(t)\bigg)\nonumber \\
&-\sum_{i=1}^I\sum_{j=1}^M R_{ij}(t)\bigg(2Q_{i,m}^\text{Sat}(t)+Q_{i,e}^\text{Sat}(t)+Z_{i,m}^\text{Sat}(t)+ A_{i,e}(t)\bigg)\nonumber \\
&\mathrm{s.t.} \quad C1-C4 \nonumber
\end{align}
 \end{footnotesize} 
%

\bibliographystyle{abbrv}
\bibliography{main}

\begin{wrapfigure}{l}{25mm} 
    \includegraphics[width=5in,height=1.25in,clip,keepaspectratio]{authors/wiem.pdf}
  \end{wrapfigure}\par
  \textbf{Wiem Abderrahim} (S'14 - M'18) accomplished her undergraduate studies in electrical engineering at the Higher School of Communications of Tunis, Carthage University, Tunisia in 2013. She received her Doctoral Degree in Information and Communication Technologies from the same university in 2017. She worked as a lecturer and then as an adjunct professor at the Higher School of Communications of Tunis between 2014 and 2018. Currently, she is a postdoctoral fellow within King Abdullah University of Science and Technology (KAUST), Thuwal, Saudi Arabia. Her research interests include cloud computing,  network virtualization and recently satellite communications and machine learning.  \\
  \begin{wrapfigure}{l}{25mm} 
    \includegraphics[width=5in,height=1.25in,clip,keepaspectratio]{authors/osama.pdf}
  \end{wrapfigure}\par
  \textbf{Osama Amin} (S'07, M'11, SM'15) received his B.Sc. degree in electrical and electronic engineering from Aswan University, Egypt, in 2000, his M.Sc. degree in electrical and electronic engineering from Assiut University, Egypt, in 2004, and his Ph.D. degree in electrical and computer engineering, University of Waterloo, Canada, in 2010. In June 2012, he joined Assiut University as an assistant professor in the Electrical and Electronics Engineering Department. Currently, he is a research scientist in the CEMSE Division at KAUST, Thuwal, Makkah Province, Saudi Arabia. His general research interests lie in communication systems and signal processing for communications with special emphasis on wireless applications.  \\
 
 \begin{wrapfigure}{l}{25mm} 
    \includegraphics[width=5in,height=1.25in,clip,keepaspectratio]{authors/slim.pdf}
  \end{wrapfigure}\par  \textbf{Mohamed Slim Alouini} (F'09) was born in Tunis, Tunisia. He received the Ph.D. degree in Electrical Engineering
from the California Institute of Technology (Caltech), Pasadena, 
CA, USA, in 1998. He served as a faculty member in the University of Minnesota,
Minneapolis, MN, USA, then in the Texas A{\&}M University at Qatar,
Education City, Doha, Qatar before joining King Abdullah University of
Science and Technology (KAUST), Thuwal, Makkah Province, Saudi
Arabia as a Professor of Electrical Engineering in 2009. His current
research interests include the modeling, design, and
performance analysis of wireless communication systems.  \\

\begin{wrapfigure}{l}{28mm} 
    \includegraphics[width=5in,height=1.25in,clip,keepaspectratio]{authors/basem.pdf}
  \end{wrapfigure}\par
  \textbf{Basem Shihada} (SM'12)  is an associate and founding professor in the Computer, Electrical and Mathematical Sciences and Engineering (CEMSE) Division at King Abdullah University of Science and Technology (KAUST). He obtained his PhD in Computer Science from University of Waterloo. In 2009, he was appointed as visiting faculty in the Department of Computer Science, Stanford University. In 2012, he was elevated to the rank of Senior Member of IEEE. His current research covers a range of topics in energy and resource allocation in wired and wireless networks, software defined networking, internet of things, data networks, smart systems, network security, and cloud/fog computing.

\end{document}